\documentclass[12pt]{article}
\usepackage{color}
\usepackage{amsmath}
\usepackage{amssymb} 
\usepackage{bm}
\usepackage[dvipdfm]{graphicx} 
 
\newcommand{\be}{\begin{equation}}
\newcommand{\ee}{\end{equation}}
\newcommand{\n}{\nonumber \\}

 \makeatletter
\def\alt{\mathrel{\mathpalette\gl@align<}}
\def\agt{\mathrel{\mathpalette\gl@align>}}
\def\gl@align#1#2{ \lower.6ex\vbox{\baselineskip\z@skip\lineskip\z@
\ialign{ $\m@th#1\hfil##\hfil$\crcr#2\crcr\sim\crcr }} } \makeatother

\begin{document}
\begin{flushright}
KEK-TH-1373
\end{flushright}
\vspace*{1.0cm}

\begin{center}
\baselineskip 20pt 
{\Large\bf 
Stochastic Analysis of an Accelerated Charged Particle \\
- Transverse Fluctuations - 
}
\vspace{1cm}

{\large 
Satoshi Iso\footnote{satoshi.iso@kek.jp} , Yasuhiro Yamamoto
\footnote{yamayasu@post.kek.jp} and Sen Zhang\footnote{zhangsen@post.kek.jp} \\

} \vspace{.5cm}

{\baselineskip 20pt \it
KEK Theory Center, Institute of Particle and Nuclear Studies, \\ 
High Energy Accelerator Research Organization(KEK)  \\
and \\
Department of Particles and Nuclear Physics, \\
The Graduate University for Advanced Studies (SOKENDAI), 
\\
1-1 Oho, Tsukuba, Ibaraki 305-0801, Japan} 

\vspace{2cm} 
{\bf Abstract} 
\end{center}
\noindent

An accelerated particle sees the Minkowski vacuum as thermally excited,
and the particle moves stochastically due to an interaction
with the thermal bath. This interaction fluctuates the particle's 
transverse momenta like
the Brownian motion in a heat bath.
Because of this fluctuating motion, 
it has been discussed that the accelerated charged particle
emits extra  radiation (the Unruh radiation  \cite{ChenTajima}) 
in addition to the classical Larmor radiation, and
 experiments are under planning to detect such radiation 
by using ultrahigh intensity lasers  constructed 
in near future \cite{ELI,ELIhp}.  
There are, however, counterarguments
that the radiation is canceled by an interference effect
between the vacuum fluctuation and the fluctuating  motion.
In fact,  in the case of an internal detector where the Heisenberg
equation of motion can be solved exactly, there is no additional radiation
after the thermalization is completed \cite{Sciama,RavalHuAnglin}.
In this paper, we revisit the issue in the case of an accelerated 
charged particle in the scalar-field analog of QED.
We  prove the equipartition theorem of transverse momenta 
by investigating a stochastic motion of the particle, and 
show that the Unruh radiation is  partially canceled
by an interference effect.

\thispagestyle{empty}
 
\newpage

\addtocounter{page}{-1}

\baselineskip 18pt
\section{Introduction}{{{
Quantum field theories in the space-time with horizons exhibit  
interesting thermodynamic behavior. The most prominent phenomenon
is the Hawking radiation \cite{HawkingRadiation} 
and the fundamental laws of thermodynamics hold
in the black hole background \cite{BHthermo}.
 This indicates an underlying microscopic
description of the space-time and there are varieties of proposals
including those based on D-branes \cite{David:2002wn} 
or quantum spin foam \cite{spinfoam}.
For a black hole with
surface gravity  $\kappa$ at the horizon, the temperature of 
the Hawking radiation is given by
\be 
T_H = \frac{\hbar \kappa}{2 \pi c k_B} =6 \times 10^{-8} 
\left( \frac{M_\odot}{M} \right) [\text{K}]  .
\ee
where we have used $\kappa = 1/4M$ for the Schwarzschild black hole with 
mass $M$.
The temperature is too small to be observed for astrophysical 
black holes. 
A similar phenomenon occurs for a uniformly accelerated observer
in the  Minkowski vacuum \cite{Unruh,Unruhrev}.
The equivalence principle of the general relativity relates 
acceleration with gravity.
If a particle is uniformly accelerated with an acceleration $a$, 
there appears a causal
horizon, the Rindler horizon, and no information can be transmitted
from the other side of the horizon.
Because of the existence of the
Rindler horizon, the accelerated observer sees
the Minkowski vacuum as thermally excited with the Unruh temperature
\be
 T_U = \frac{\hbar a}{2 \pi c k_B} = 4 \times 10^{-23}
  \left( \frac{a}{1 \,\text{cm/s}^2} \right)[\text{K}].
\ee
Furthermore it is discussed \cite{Jacobson} that,
if we assign entropy to the  Rindler horizon and assume thermodynamic relations,
the Einstein equation  can be derived.

The Unruh temperature is very small 
for  low acceleration, but the recent development of 
ultra-high intensity lasers makes the Unruh effect experimentally accessible.
In the electro-magnetic field of a laser with intensity $I$[W/cm$^2$], 
an electron can be accelerated to
\be
 a = 2 \times 10^{12} \ \sqrt{\frac{I}{1\,\text{W/cm}^2}}\ [\text{cm/s}^2]
\ee
and the Unruh temperature is given by
\be
 T_U = 8 \times 10^{-11} \sqrt{\frac{I}{1\,\text{W/cm}^2}}\ [\text{K}].
\ee
The ELI (Extreme Light Infrastructure) project \cite{ELIhp}
recently approved is planning to construct  Peta Watt lasers with an intensity
as high as $5 \times 10^{26} \ [\text{W/cm}^2]$. 
Then the expected Unruh temperature becomes
more than $10^3 \,\text{K}$ which is
 much higher than the room temperature. 
Can we experimentally observe such  high Unruh temperature
of an accelerated electron in the laser field?
This is an interesting issue and worth being investigated.

One such proposal was given by P.Chen and T.Tajima 
\cite{ChenTajima}. Their basic idea is the following. 
An electron is accelerated in the oscillating electro-magnetic
field of lasers. It is not a uniform acceleration, but
they approximated the electron's motion around the 
turning points by a uniform acceleration. 
Since the electron feels the vacuum as thermally excited
with the Unruh temperature, the motion of the 
electron will be thermalized  and  fluctuate around the classical trajectory
(Fig.~\ref{Fluctuation}).
\begin{figure}[htb]
\begin{center}
  \includegraphics[width=10em]{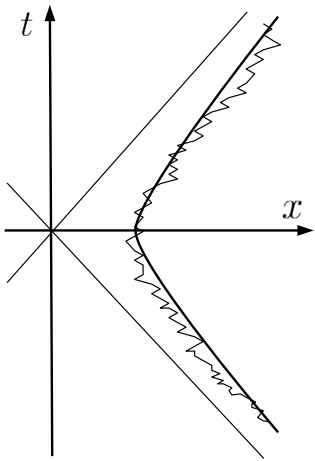}
 \caption{Stochastic trajectories induced by quantum field fluctuations.} 
 \label{Fluctuation}
\end{center}
\end{figure}
Because of this fluctuating motion of an electron,
they conjectured that 
 additional radiation, apart from the classical Larmor
radiation, will emanate.  
Using an intuitive argument, they estimated the additional radiation
and  called it Unruh radiation. 
Though the estimated amount of radiation is much smaller than
the classical one by $ 10^{-5}$,
they argued that the angular dependence is different.
Especially there is a blind
spot for the Larmor radiation in the direction along the acceleration  
while the Unruh radiation is expected to be radiated more spherically. 
Hence they proposed to detect the additional radiation in 
this direction. 

The above heuristic argument sounds correct, but
it has been known in a  simpler situation
that such  radiation is canceled by 
an interference effect between the radiation field emanated 
from the fluctuating motion and the vacuum fluctuation 
of the radiation field  \cite{Sciama,RavalHuAnglin}.
The cancellation was shown to 
occur\footnote{ We note that, though the radiation vanishes in generic points,
 a singular behavior of the radiation
is pointed out to exist on the past horizon \cite{RefUnruhSingular}.} for an internal detector 
in 1+1 dimensions\footnote{
We call a detector with internal degrees of freedom an internal detector
in order to distinguish it from an accelerated charged particle. In the case of a charged
particle, the position of the particle reacts to the thermal effect of acceleration.
On the contrary, in the case of an internal detector, only its internal degree of freedom
is excited by the thermal effect.
}.
In order to see the importance of the interference effect, we briefly 
sketch the calculation of radiation from the internal detector.
In these papers \cite{Sciama,RavalHuAnglin}, 
the authors analyzed 
a uniformly accelerated internal detector $Q$ coupled with a scalar
field $\phi$ in 1+1 dimensions. 
The action is given by
\be
S = S(Q) + S(\phi) + e \int d\tau \frac{dQ}{d\tau} \phi(z(\tau))
\ee
where $z(\tau)=(t(\tau),x(\tau))$ represents 
a classical trajectory of the detector. $S(Q)$ and $S(\phi)$ are quadratic  actions of 
the internal degrees of freedom of the detector
(i.e. a harmonic oscillator) and the scalar field in 1+1 dimensions
respectively;
\begin{align}
 S(Q) &= 
  \int d\tau \left( \frac{1}{2} \dot{Q}(\tau)^2 
                   -\frac{\omega_0^2}{2} Q^2 (\tau) \right) \\
 S(\phi) &=
  \int d^2x\, \frac{1}{2} (\partial \phi (x))^2 .
\end{align}
Since the coupling term is linear both in $Q$ and $\phi$,
the Heisenberg equations of motion can be exactly solved.
A classical solution of the scalar  field is written as a sum of the
vacuum fluctuation $\phi_h(x)$,  
which is a solution to the homogeneous equation in the absence of $Q$,
and an inhomogeneous term $\phi_{inh}(x)$  as
\begin{equation}
\phi(x) = \phi_h(x) + \phi_{inh}(x).
\end{equation}
The inhomogeneous part $\phi_{inh}$ is given by 
\begin{equation}
 \phi_{inh}(x) = \int d\tau \, G_R(x,z(\tau)) \frac{dQ}{d\tau} 
 \label{phir}
\end{equation}
where $G_R$ is the retarded Green function of the scalar field. 
The equation of motion of $Q$ becomes
\be
 \ddot{Q} + \omega_0^2 Q  = -e \frac{d\phi_h}{d\tau} 
 - e \frac{d\phi_{inh}}{d\tau}.
\label{EOMdet}
\ee
Since $\phi_{inh}$ is solved linearly in $Q(\tau)$.,
the second term of the r.h.s. of (\ref{EOMdet}) is linear in $Q$.
It can be shown that it becomes a dissipative term  
$\gamma \dot{Q}$  
where $\gamma = e/2\pi$.
Hence $Q(\tau)$ can be solved in terms of the homogeneous solution $\phi_h$ as 
\be
 \tilde{Q}(\omega) = e h(\omega) \varphi(\omega)
\label{111}
\ee
where $\tilde{Q}(\omega)$  and $\varphi(\omega)$
are the Fourier modes of $Q(\tau)$ and $\phi_h(z(\tau))$
with respect to $\tau$, and 
$h(\omega) = i \omega/(\omega^2 -\omega_0^2 -i\omega \gamma)$.
By inserting  (\ref{111}) to (\ref{phir}), the inhomogeneous solution 
$\phi_{inh}$ is solved in terms 
 of the vacuum fluctuation $\phi_{h}(x)$.

Then it is straightforward to calculate the energy-momentum tensor. 
Since the energy flux is written in terms of the 2-point function,
they first calculated the 2-point function 
\be
G(x,x') =\langle \phi(x) \phi(x') \rangle.
\ee
It is written as a sum of the following terms,
\begin{align}
G(x,x') - G_0(x,x') &=
\langle \phi_{inh}(x) \phi_{inh}(x') \rangle
+ \langle \phi_{inh}(x) \phi_h(x') \rangle
+ \langle \phi_h(x) \phi_{inh}(x') \rangle
\end{align}
where  vacuum fluctuation
$G_0(x,x') = \langle \phi_h(x) \phi_h(x') \rangle$
is subtracted.
The first term $\langle \phi_{inh}(x) \phi_{inh}(x') \rangle$
can be considered as 
an analog of the Unruh radiation
proposed in \cite{ChenTajima}.
It is  nonzero because the  detector is 
thermally excited from the classical ground state $Q=0$.
However, Sciama et.al. \cite{Sciama} and Hu. et.al. \cite{RavalHuAnglin}
have shown that in (1+1)-dimensional case, the contributions from the interference terms 
$\langle \phi_{inh}(x) \phi_h(x') \rangle
+ \langle \phi_h(x) \phi_{inh}(x') \rangle$ 
cancel the radiation
$\langle \phi_{inh}(x) \phi_{inh}(x') \rangle$,
except for the polarization cloud.
The polarization cloud is the cloud of radiation field localized near the
accelerated particle, and
does not contribute to the energy-momentum tensor.
Since we are interested in the energy flux far from the particle in this paper,
we do not consider it in the following.
Hu and Lin \cite{HuLin}
extended the calculation to a detector with internal degrees of freedom
in 3+1 dimensions. The calculation in the 3+1 dimensional case 
is  reviewed in Appendix A.

The physical reason of the cancellation is discussed  in \cite{Grove:1986} and \cite{Massar:2005vg} (see also \cite{Brout:1995rd}).
The key point of their arguments is
 the invariance of the Minkowski vacuum state under boosts with the conservation of energy-momentum tensor.
The metric of the Minkowski space is invariant under the translation of $\tau$.
So, after the thermalization occurs,
the system becomes stationary and 
the energy-momentum tensor will not depend on $\tau$ explicitly.
In  \cite{Massar:2005vg}, 
the following three conditions are assumed:
\begin{itemize}
\item  Poincare invariance;
\item A detector follows a uniformly accelerated trajectory in flat space;
\item A $\tau$-independent coupling.
\end{itemize}
With these conditions,  the energy-momentum tensor does not depend 
explicitly on $\tau$. They have argued that,
if the energy conservation holds for the total system,
there is no outgoing flux.
Namely the cancellation of the radiation
$\langle \phi_{inh}(x) \phi_{inh}(x') \rangle$
by the interference terms is not accidental but a consequence of the 
boost symmetry.
In the case of an accelerated charged particle, however,
the second condition does not hold since 
the particle's trajectory is a dynamical variable and fluctuates.
Then the boost invariance of the trajectory is lost.
Furthermore, the charged particle is accelerated by injecting energy
from outside.
It is therefore not obvious whether extra radiation vanishes or not.

The purpose of the paper is to investigate
a stochastic motion of a uniformly accelerated charged particle 
and to study whether there is  additional radiation, Unruh radiation  
associated with the stochastic motion of the particle.
The situation becomes much more complicated than the internal detector
case because 
the equations of motion are highly nonlinear.
When the particle's motion $z(\tau)$ is affected by the vacuum fluctuation, 
the Green function $G_R(x, z(\tau))$ is also changed accordingly unlike the internal detector case. 
Hence  in order to calculate the radiation from an accelerated charged particle,
we need to approximate the fluctuating motion near the classical trajectory
and assume  that fluctuations are small.
We first study a stochastic equation for a charged particle
coupled with the scalar field \cite{JohnsonHu}. 
This gives a simplified model of the real QED. 
A self-interaction with the scalar field
created by the particle itself gives a backreaction to the 
particle's motion, and it gives  a
radiation damping term of the Abraham-Lorentz-Dirac equation \cite{ALD}.
If we further regard the vacuum fluctuation as  stochastic noise, 
the particle's motion obeys a generalized Langevin equation.
By solving the Langevin equation,
we can obtain  stochastic fluctuations of the particle's momenta. 
In this paper, we mainly focus on fluctuations in the transverse directions
and leave analysis of longitudinal fluctuations for future investigation.
Some comments are given in Discussions and in Appendix \ref{app2}.

The organization of the paper is the following.
In Sec.~II, we summarize the basic framework of our system, and
obtain a generalized Langevin equation for a charged particle
coupled with a scalar field.
In Sec.~III, we consider small fluctuations in the transverse directions.
Then the stochastic equation can be solved and we can prove the
equipartition theorem for the transverse momenta, 
i.e., a stochastic average of
a square of  momentum fluctuations in the
transverse directions is shown to be proportional to the Unruh 
temperature. 
We also discuss the relaxation time of the thermalization process. 
Section~IV is the main part of the paper.
We calculate radiation emitted by a charged particle in the scalar-field analog of QED.
The interference terms partially cancel the radiation coming from
the contribution 
$\langle \phi_{inh}(x) \phi_{inh}(x')\rangle$, but
unlike the case of an internal detector, they do not cancel exactly. 
In Sec.~V, we obtain a similar stochastic equation for
an accelerated charged particle in the real QED, and show
the equipartition theorem for transverse momenta. 
Section~VI is devoted to conclusions and discussions.
In Appendix A, we review the calculation of radiation
for a uniformly accelerated internal detector 
in (3+1) dimensions~\cite{HuLin}.
In Appendix \ref{app2}, we consider  fluctuations in the 
longitudinal and temporal directions. 
We use Planck units $c=\hbar=k_B=1$ in most of the paper.
\section{Stochastic equation of an accelerated particle} 
\setcounter{equation}{0}
We consider the scalar-field analog of QED. The model is analyzed in~\cite{JohnsonHu}
and here we briefly review the settings and the derivation of
the stochastic Abraham-Lorentz-Dirac (ALD) equations. In~\cite{JohnsonHu}, the authors
used the influence functional approach, but here we take a simplified method. 
The system composes of a relativistic particle $z^\mu(\tau)$ and
the scalar  field $\phi(x)$. The action is given by
\begin{align}
 S[z,\phi, h] = S[z,h] + S[\phi] + S[z,\phi],
\end{align}
with
\begin{align}
 S[z,h] &= 
   -m \int d\tau \  \sqrt{\dot{z}^{\mu} \dot{z}_\mu}, \\
 S[\phi] &= \int d^4x \ \frac{1}{2}(\partial_\mu \phi (x))^2 , \\
 S[z,\phi] &= \int d^4x \ j(x;z) \phi(x) .
\end{align}
The scalar current $j(x;z)$ is defined as
\begin{align}
 j(x;z) = e \int d\tau \ \sqrt{\dot{z}^\mu \dot{z}_\mu}\, \delta^4 (x-z(\tau)),
 \label{current}
\end{align}
where $e$ is negative for an electron.
We can parametrize the particle's path satisfying  
 $\dot{z}^2=1$ by taking $\tau$ properly.
 
The equation of motion of the particle is given by
\begin{equation}
 m\ddot{z}^\mu = 
   F^\mu - \int d^4 x \frac{\delta j(x;z)}{\delta z_\mu (\tau)} \phi(x)
\label{particle-eom}
\end{equation}
where we have added the external force $F^\mu$ so as to
 accelerate the particle uniformly:
\begin{equation}
 F^\mu = ma(\dot{z}^1, \dot{z}^0, 0,0).
\end{equation}
Then a classical solution of the particle (in the absence of
 the coupling to $\phi$) is given by
\begin{equation}
  z^\mu_{\textrm{cl}} = (\frac{1}{a}\sinh{a\tau},\frac{1}{a}\cosh{a \tau},0,0).
 \label{classical-trajectory}
\end{equation}
Note that the external force satisfies $F^\mu \dot{z}_\mu =0$ and therefore
the classical equation of motion preserves the gauge condition $\dot{z}^2=1.$
From the definition of the current (\ref{current}), it is easy to prove
the identity,
\begin{equation}
 \int d^4 x \frac{\delta j(x;z)}{\delta z^\mu (\tau)} f(x) = 
   e \overrightarrow{\omega}_\mu f(x)|_{x=z(\tau)}
\end{equation}
where $\overrightarrow{\omega}^\mu$ is given by
\begin{align}
 \overrightarrow{\omega}_\mu = \dot{z}^\nu \dot{z}_{[\nu}
  \partial_{\mu]} - \ddot{z}_\mu.
\end{align}
Here we have used the gauge condition $\dot{z}^2=1$ and 
$\ddot{z} \cdot \dot{z}=0$.
Hence the equation of motion (\ref{particle-eom}) becomes
\begin{equation}
 m\ddot{z}^\mu = F^\mu - e \overrightarrow{\omega}^\mu \phi(z(\tau))
\label{particle-eom2}
\end{equation}
Since the differential operator $\overrightarrow{\omega}_\mu$ satisfies
$\dot{z}^\mu  \overrightarrow{\omega}_\mu =0$ for a classical path satisfying
the gauge condition,
the stochastic Eq.~(\ref{particle-eom2}) continues to preserve the 
 condition $\dot{z}^2=1$. The second term of (\ref{particle-eom2})
 represents a self-interaction of the particle with the radiation 
 emitted by the particle itself.

The equation of motion of the radiation field 
$ \partial^\mu \partial_\mu \phi(x) = j(x) $
is  solved by using the retarded Green function 
$G_R$ as
\begin{equation}
 \phi(x) = \phi_{h} (x) + \phi_{inh}, \qquad
 \phi_{inh}= \int d^4 x' G_R(x,x') j(x';z) 
 \label{sol-phi}
\end{equation}
where $\phi_h$ is the homogeneous solution of the equation of motion and 
represents the vacuum fluctuation.
The retarded Green function satisfies
\begin{equation}
 \partial^\mu \partial_\mu G_R (x,x') = \delta^{(4)} (x-x')
\end{equation}
and is given by
\begin{align}
 G_R(x,x') &= 
 i \langle [\phi(x), \phi(x')] \rangle \theta(t-t') \\ &= 
   \frac{\theta(t-t') \delta((x-x')^2)}{2\pi} =\frac{\delta((t-t')-r)}{4\pi r}
\label{RG}
\end{align}
where $r^2=|{\bf x}-{\bf x'}|^2$. 
Inserting the solution (\ref{sol-phi}) into (\ref{particle-eom2}),
we have the following stochastic equation for the particle
\be
 m\ddot{z}^\mu(\tau) = F^\mu(z(\tau)) - e \overrightarrow{\omega}^\mu 
 \left( \phi_h(z(\tau)) + e \int d\tau' \ G_R(z(\tau), z(\tau'))  \right) .
\label{stochastic}
\ee
Here we have used the gauge condition $\dot{z}^2=1$.
The operator $\overrightarrow{\omega}_\mu$ acts on $z(\tau)$.
The homogeneous part $\phi_h(z(\tau))$ of the scalar field
describes  Gaussian
fluctuations of the vacuum, and hence the first term in the
parenthesis can be interpreted as random noise to the particle's motion. 
Expanding $\phi_h$  as
\begin{equation}
 \phi_h (x) = \int \frac{d^3 k}{(2\pi)^3} \frac{1}{\sqrt{2 \omega_k}}
 (a_k e^{-i k^\mu x_\mu} + a_k^\dagger e^{i k^\mu x_\mu}),
\end{equation} 
the vacuum fluctuation is given by
\begin{equation}
 \langle \phi_h(x) \phi_h (x') \rangle 
 = - \frac{1}{4 \pi^2}  \frac{1}{(t-t'-i\epsilon)^2-r^2}.
\label{phi-correlation} 
\end{equation} 
It is essentially quantum mechanical,  but if it is 
evaluated on a world line
 of a uniformly accelerated particle $x=z(\tau), x'=z(\tau')$,
it behaves as  the ordinary finite temperature noise. 

The second term in the parenthesis of (\ref{stochastic})
is a functional of the total history of the particle's
motion $z(\tau')$ for $\tau' \le \tau$, 
but it can be reduced to
 the so called radiation damping term
 of a charged particle coupled with a radiation field.
It is generally nonlocal, but since the Green function damps
rapidly as a function of the distance $r$, the term 
is approximated by local derivative terms.  
First define $y^\mu(s)=z^\mu(\tau) - z^\mu(\tau')$ where
$\tau^\prime=\tau -s$ with $\tau$ kept fixed. Then
it can be expanded as
\begin{equation}
y^{\mu}(s) = s \dot{z}^\mu(\tau) -\frac{s^2}{2} \ddot{z}^\mu(\tau) 
+ \frac{s^3}{6} \dddot{z}^\mu (\tau) + \cdots.
\end{equation}
A square of the space-time distance $\sigma$ is given by
\begin{align}
 \sigma(s) \equiv  y^\mu y_\mu 
 = s^2 \left(1-\frac{s^2}{12}(\ddot{z})^2 +\cdots \right)
\end{align}
and
\begin{equation}
\frac{d\sigma(s)}{ds} = 2 y^\mu \dot{y}_\mu 
= 2s \left(1-\frac{s^2}{6}(\ddot{z})^2 +\cdots \right).
\end{equation}
In deriving them, we have used the gauge condition 
$(\dot{z})^2=1$, $\dot{z} \cdot \ddot{z}=0$ and $\dot{z} \cdot \dddot{z}=-(\ddot{z})^2$.
The derivative $\partial_\mu$ appearing in the operator 
$\overrightarrow{\omega}_\mu$ can be written
in terms of  $\frac{d}{ds}$,  when it acts on a function of $\sigma$, 
 as
\begin{align}
 \partial_\mu = 
 \frac{\partial \sigma}{\partial z^\mu} \frac{d}{d\sigma}
  &= 2y_\mu \left(\frac{d\sigma}{ds} \right)^{-1} \frac{d}{ds} =
  \frac{y_\mu}{y^\nu \dot{y}_\nu} \frac{d}{ds} 
  \\
  &= \left(\dot{z}_\mu - \frac{s}{2}\ddot{z}_\mu +
\frac{s^2}{6} (\dddot{z}_\mu+\dot{z}_\mu (\ddot{z})^2)+\cdots\right) 
\frac{d}{ds}.
\end{align}
Hence the second term in the parenthesis of (\ref{stochastic}) can be
simplified as 
\begin{align}
 &e^2 \overrightarrow{\omega}_\mu \int_{-\infty}^{\tau} d\tau' \,
 G_R (z(\tau),z(\tau')) \\
 =& 
 e^2 \int_0^\infty ds\, \overrightarrow{\omega}_\mu G_R(s) 
 \\
 =& e^2 \int^{\infty}_0 ds  \biggl( a_\mu(\tau)G^R(s) + a_\mu(\tau) \frac{s}{2}
  \frac{d}{ds} G^R(s) 
+ (\dot{z}_\mu \ddot{z}^2 + \dddot{z}_\mu) \frac{s^2}{6} \frac{d}{ds}
 G^R(s) + \mathcal{O}(s^3) \biggr).
\end{align}
In the first equality, we have neglected the singular term proportional
to $\delta(\sigma)$. 
The first two terms can be absorbed by  mass renormalization. 
The $s$ integrals of them are divergent, 
so we assume that the renormalized mass becomes finite
after the mass renormalization.
The last
one is the radiation reaction term and can be evaluated by using the identity
\begin{equation}
  \int^{\infty}_0 ds\, s^2 \frac{d}{ds} G^R(s)
  = \int^{\infty}_0 ds\, s^2 \frac{d}{ds} \frac{\delta(s)}{4 \pi s}
  = - \frac{1}{2 \pi}
\label{EqRetexp}
\end{equation}

After the mass renormalization, we
get the following generalized Langevin equation for the charged particle,
\begin{equation}
 m \ddot{z}^\mu 
 - F^\mu - \frac{e^2}{12\pi} (\dot{z}^\mu \ddot{z}^2 + \dddot{z}^\mu) 
  =
 - e \overrightarrow{\omega}^\mu \phi_h(z).
\label{stochastic-particle}
\end{equation}
This is an analog of the Abraham-Lorentz-Dirac equation for a charged particle
interacting with the electromagnetic field.
The dissipation term is induced through an effect of 
 backreaction of the particle's radiation to the particle's motion. 
Note that, if the noise term, the r.h.s.,  is absent, 
the classical solution (\ref{classical-trajectory}) 
is still a solution to Eq.~(\ref{stochastic-particle}).
\section{Thermalization of transverse momentum fluctuations} 
\setcounter{equation}{0}
The stochastic Eq.~(\ref{stochastic-particle}) is 
nonlinear and difficult to solve. Here we 
consider small fluctuations around the
classical trajectory induced by the
vacuum fluctuation $\phi_h$. 
Especially, we consider fluctuations
in the transverse directions perpendicular to the direction of the acceleration.
The classical trajectory of the particle does not move to the transverse directions,
and we can easily treat fluctuations in these directions. 
In contrast, the particle is accelerated strongly to the longitudinal direction
and it is more difficult to separate fluctuations from the classical solution.
 We shortly discuss longitudinal fluctuations in Appendix \ref{app2}.

First we expand the particle's motion around the classical trajectory 
$z^\mu_0$ as
\begin{equation}
 z^\mu(\tau) = z^\mu_0 + \delta z^\mu.
\end{equation} 
The particle is accelerated  along the $x$ direction.
Now we consider small fluctuation in transverse directions $z^i$.
By expanding the stochastic Eq.~(\ref{stochastic-particle}), 
we can obtain a linearized stochastic equation for
the transverse velocity fluctuation $\delta v^i \equiv \delta \dot{z}^i$ as,
\begin{equation}
 m \delta \dot{v}^i = 
 e \partial_i \phi_h + \frac{e^2}{12\pi} 
(\delta \ddot{v}^i -a^2 \delta v^i) .
\label{stochastic-transverse}
\end{equation}
Performing   Fourier transformation with respect to the trajectory's 
parameter
$\tau$
\begin{equation}
 \delta v^i(\tau) = \int \frac{d \omega}{2\pi} 
 \delta \tilde{v}^i(\omega) e^{-i \omega \tau}, \ \ 
 \partial_i \phi_h(\tau) = 
   \int \frac{d \omega}{2\pi} \partial_i \varphi(\omega) 
	     e^{-i \omega \tau},
\label{varphi}
\end{equation}
the stochastic equation can be solved as
\begin{equation}
 \delta \tilde{v}^i(\omega) = 
 e h(\omega) \partial_i \varphi(\omega)
\end{equation}
where 
\begin{equation}
 h(\omega) = \frac{1}{-im \omega +\frac{e^2(\omega^2 +a^2)}{12\pi}}.
\end{equation}

The vacuum 2-point function on the classical 
trajectory can be evaluated  from (\ref{phi-correlation}) as
\begin{align}
 \langle \partial_i \phi_h(x) \partial_j \phi_h (x') \rangle 
 |_{x=z(\tau),x'=z(\tau')}
 =& \frac{1}{2\pi^2} 
     \frac{\delta_{ij}}{((t-t'-i \epsilon)^2 -r^2)^2} \\
 =& \frac{a^4}{32 \pi^2}  
     \frac{\delta_{ij}}{\sinh^4 \left(\frac{a(\tau-\tau'-i \epsilon)}{2}\right)}.
\label{phi-correlation2} 
\end{align}
It has originated from the quantum fluctuations of the vacuum,
but it can be interpreted as finite temperature noise if it is 
evaluated on the accelerated
particle's trajectory \cite{Unruh}
\footnote{The finite temperature (Unruh) effect 
is caused by the appearance of the horizon for a uniformly 
accelerated observer in the Minkowski space-time
and  analogous to the Hawking radiation of the black hole,
but you should not confuse
the radiation we are discussing in this paper with
the Hawking radiation. The accelerated observer
sees the Minkowski vacuum as thermally excited, but it is excited 
from the Rindler vacuum (not from the Minkowski vacuum) 
and the energy-momentum tensor remains zero as ever. 
The radiation discussed in the paper is, if exists, produced
by an interaction with the vacuum and the accelerated charged particle.}.
By making the Fourier transformation with respect to $\tau$,
we have
\begin{equation}
 \langle \partial_i \varphi(\omega) \partial_j 
 \varphi(\omega') \rangle 
= 2 \pi \delta(\omega+\omega') \delta_{ij} I(\omega)
\end{equation}
where
\begin{align}
 I (\omega)= \frac{a^4}{32 \pi^2} \int_{-\infty}^{\infty} d\tau
 \frac{ e^{i \omega \tau} }{ \sinh^4(\frac{a(\tau-i \epsilon)}{2}) } 
 =& \frac{1}{6 \pi} \frac{\omega^3 + \omega a^2}{1-e^{-2\pi \omega/a}} .
\end{align}
For small $\omega$, this is expanded as
\begin{align}
I(\omega)=\frac{a}{12\pi^2} (a^2 + a\pi \omega 
 + \cdots).
\end{align}
The expansion corresponds to the derivative
expansion 
\begin{align}
 \langle \partial_i \phi_h(x) \partial_j \phi_h (x') \rangle 
 |_{x=z(\tau),x'=z(\tau')}
 &= \frac{a^3}{12\pi^2} \delta_{ij} \delta (\tau-\tau') 
  -i \frac{a^2}{12\pi  } \delta_{ij} \delta'(\tau-\tau')
 +\cdots .
\end{align}
If we approximate the 2-point function by the first term,
the noise correlation becomes  white noise. 
The coefficient  determines
the strength of the noise.  We show that it is consistent
with the fluctuation-dissipation theorem at the Unruh temperature. 

By symmetrizing it, i.e.,
$\langle \partial \phi(x) \partial \phi(x')\rangle_S
= \langle \{\partial \phi(x), \partial \phi(x') \} \rangle /2$,
the correlation function becomes
\begin{align}
 I_S (\omega)= 
 \frac{\omega (\omega^2 + a^2)}{12 \pi} 
   \coth \left(\frac{\pi \omega}{a} \right).
\end{align}
It is an even function of $\omega.$ The correlators $I(\omega)$
and $I_S(\omega)$ should be regularized for large $\omega$ 
where quantum field theoretic effects of electrons become
important. 

The expectation value of a square of the 
velocity fluctuations in the transverse directions can be evaluated as
\begin{align}
  \langle
    \delta v^i(\tau) \delta v^j(\tau^\prime)
  \rangle_S
  &= 
  e^2 \int \frac{d \omega d\omega'}{(2\pi)^2} 
  \langle
    \partial_i \varphi(\omega) \partial_j \varphi(\omega')
  \rangle_S \ 
  h (\omega) h (\omega') e^{-i (\omega \tau+\omega' \tau')}
  \\
  &= 
  e^2 \delta_{ij} \int \frac{d \omega }{2\pi} I_S  (\omega) |h(\omega)|^2 
  e^{-i \omega (\tau -\tau')} 
  \\
  &\sim 
  e^2 \delta_{ij} \int \frac{d \omega }{24 \pi^3} 
  \frac{a^3}
       {(m \omega)^2+\left(\frac{e^2}{12\pi}\right)(\omega^2+a^2)^2} 
  e^{-i \omega (\tau -\tau')} .
\label{momentum-fluctuation}
\end{align} 
The denominator has four poles at $\omega=\pm i \Omega_{\pm}$
where
\begin{align}
  \Omega_+ &= \frac{12\pi m}{e^2}(1+{\cal O}(a^2/m^2)), \\
  \Omega_- &= \frac{a^2 e^2 }{12\pi m}(1+{\cal O}(a^2/m^2)).
\end{align} 
The acceleration of an electron in high-intensity laser fields
in the near future
can be at most 0.1 eV and much smaller than the electron mass 0.5 MeV. Hence, 
the values of these poles satisfy the following inequalities,
\begin{equation}
 \Omega_+ \gg a \gg \Omega_- .
\end{equation}
Since the energy scale of the dynamics of the accelerated particle
is much smaller than the electron mass, the poles at $  \pm i \Omega_+$
should be  considered  spurious 
and we should not take the contributions of the 
residues at $\pm i \Omega_+$\footnote{ 
Or we can simply approximate the denominator by dropping the
$\omega^4$ term. Then only the poles at $\pm i \Omega_-$ survive.}. 
By taking the residue at $\pm \omega= i \Omega_-$,  
we can evaluate the integral and get the following result,
\begin{equation}
\frac{m}{2}
\langle
  \delta v^i(\tau) \delta v^j(\tau)
\rangle
= \frac{1}{2}\frac{a \hbar}{2\pi c} \delta_{ij}
  \left( 1 + \mathcal{O} (a^2/m^2) \right).
\label{equipartition}
\end{equation}
Here we have recovered $c$ and $\hbar$.
This gives the equipartition relation for the transverse 
momentum fluctuations at the Unruh temperature 
$T_U=a \hbar /2\pi c$. 
The typical energy scale of the fluctuating motion is given by the
value of the pole $\Omega_-$. Since it is much smaller than the acceleration,
the derivative expansion with respect to $\omega/a$ is justified for the transverse fluctuations.

The thermalization process of the
stochastic Eq.~(\ref{stochastic-transverse}) can be also 
discussed. For simplicity,  we  approximate the stochastic
equation by dropping the 
second derivative term. 
This corresponds to the derivative expansion with respect to $\omega/a$.
Then it is solved as
\begin{equation}
 \delta v^i(\tau) = e^{-\Omega_- \tau}\delta v^i(0)
+ \frac{e}{m} \int_0^\tau d\tau^\prime \ \partial_i \phi(z(\tau')) 
e^{-\Omega_- (\tau-\tau^\prime)} .
\end{equation}
The relaxation time is given by $\tau_R=1/\Omega_-.$
The momentum square can be  calculated as
\begin{align}
 \langle
  \delta v^i(\tau) \delta v^j(\tau) 
 \rangle
  =& 
   e^{-2 \Omega_- \tau} \delta v^i(0) \delta v^j(0) \n &
  +e^2 \int_0^\tau d\tau' \ \int_0^\tau d\tau'' \ 
   e^{- \Omega_- (\tau-\tau')}e^{- \Omega_- (\tau-\tau'')}
 \langle
  \partial_i \phi(z(\tau')) \partial_j \phi(z(\tau''))
 \rangle \n
 =& e^{-2 \Omega_- \tau} \delta v^i(0) \delta v^j(0)
 + \frac{a \delta_{ij}}{2\pi m} (1- e^{-2 \Omega_- \tau}).
\end{align}
For $\tau \rightarrow \infty$, it approaches the thermalized average
(\ref{equipartition}). 
The relaxation time in the proper time
can be estimated, for $a = 0.1$ eV and $m=0.5$ MeV, to be
\begin{equation}
 \tau_R  = \frac{12\pi m}{a^2 e^2} = 1.4 \times 10^{-5} \text{sec}.
\end{equation}
This relaxation time should be compared with the laser frequency. 
The oscillation period of the laser field at ELI is 
about $3 \times 10^{-15}$ seconds and 
much shorter than the above relaxation time.
Hence an accelerated electron by the ELI laser 
does not thermalize during each oscillation
and the uniform acceleration is not a good approximation.
Even in such a situation, if an electron is 
accelerated in the laser field for a long time,
it may feel  averaged temperature. 
In order to fully understand the Unruh effect of 
nonuniform acceleration, we need to 
investigate the transient phenomena.
It is beyond the present calculation, and we leave it for future
investigations. See e.g. \cite{RefEmil} for the Unruh effect of the circular motion,
.

The position of the particle in the transverse directions 
fluctuates like the ordinary 
Brownian motion in a heat bath.
A mean square of the fluctuations is given by
\begin{align}
 R^2(\tau) =
 \sum_{i=y,z} \langle (z^i(\tau)-z^i(0))^2 \rangle = 
  2D \left( 
    \tau - \frac{3-4 e^{-\Omega_- \tau} + e^{-2\Omega_- \tau}}{2\Omega_-} 
  \right).
\end{align}
The diffusion constant $D$ is given by
\be
D = \frac{2T_U}{\Omega_- m} = \frac{12}{a e^2},
\ee
which is estimated for the above parameters
as $D \sim 7.8 \times 10^{4} \ m^2/s $.
In the Ballistic region where $\tau < \tau_R$, the mean
square becomes 
\be
R^2(\tau) = \frac{2T_U}{m} \tau^2 
\ee
 while in the diffusive region 
($\tau > \tau_R$), it is proportional to the proper time
as
\be
R^2(\tau) = 2D\tau.
\ee
As the ordinary Brownian motion, the mean square of the particle's
transverse position grows linearly with time.
If it becomes possible to accelerate the particle 
for a sufficiently long period, it may be possible 
to detect such a Brownian motion in future laser experiments.
\section{Quantum Radiation by Transverse Fluctuation} 
\setcounter{equation}{0}
Once we  obtain the stochastic motion of the accelerated particle,
it is straightforward to calculate the energy flux of the radiation 
field emitted by this particle.
In this section, we calculate the radiation induced by the fluctuation
in the transverse directions. 
\subsection{2-point function}
First we evaluate  the 2-point function
\begin{align}
G(x,x') - G_0(x,x') =&
  \langle \phi_{inh}(x) \phi_{inh}(x') \rangle
+ \langle \phi_{inh}(x) \phi_h(x')     \rangle
+ \langle \phi_h(x)     \phi_{inh}(x') \rangle.
\end{align}
The inhomogeneous solution 
$\phi_{inh}$ is produced by the accelerated charged particle
while the homogeneous solution $\phi_h$
represents the vacuum fluctuation of the quantum field $\phi$.
The Unruh radiation estimated in \cite{ChenTajima} corresponds to 
calculating the 2-point correlation function of the inhomogeneous terms 
$\langle \phi_{inh}(x) \phi_{inh}(x') \rangle$. As we will see later,
this term also contains the classical Larmor radiation.
However, this is not the end of the story.
As it has been discussed  in \cite{Sciama},
the interference terms  
$\langle \phi_{inh} \phi_h \rangle + \langle \phi_h \phi_{inh} \rangle$
cannot be neglected and 
may possibly  cancel the Unruh radiation 
in $\langle \phi_{inh} \phi_{inh} \rangle$ after the thermalization 
occurs.
This is shown for an internal detector in (1+1) dimensions,
but it is not obvious whether the same cancellation occurs for 
the case of a charged particle we are considering.

The inhomogeneous solution of the scalar field is written in terms of the 
position of the accelerated particle $z^\mu(\tau)$ as
\begin{align}
  \phi_{inh} (x)
  =& e \int d\tau G_R (x -z(\tau))  \\
  =& e \int d\tau \frac{\theta(t-z^0(\tau)) \delta((x-z(\tau))^2)}{2\pi}  
  = \frac{e}{4\pi \rho} .
  \label{trans-inh}
\end{align} 
where $\rho$ is defined by
\begin{align}
  \rho =& \dot{z}(\tau^x_{-}) \cdot (x -z(\tau^x_{-})).
  \label{rhodef}
\end{align}
Because of the step and the delta functions in the integrand of 
(\ref{trans-inh}), 
$\tau^x_-$ satisfies
\begin{align}
(x-z(\tau^x_{-}))^2 =& 0, \quad x^0 > z^0(\tau^x_{-}),
\end{align}
which is the proper time of the particle whose radiation
travels to the space-time point  $x$. 
Hence $z(\tau_-^x)$ lies on an intersection between the particle's
world line and the light-cone extending from the observer's position $x$
(see Fig. 2).
We write the superscript $x$ to make the $x$ dependence of $\tau$ 
explicitly.
The meaning of the subscript $(-)$ will be made clear later.
By using the light-cone condition, $\rho$ can be  rewritten as
\begin{align}
 \rho(x) = \frac{dz^0(\tau_-^x)}{d \tau} r(\tau_-^x) 
 \left(1- \frac{{\bf v}\cdot{\bf r}}{r}\right)
 \label{rhoLorentz}
\end{align}
where ${\bf v}=\frac{d{\bf z}}{dz^0}$, 
${\bf r}(\tau_-^x)={\bf x} -{\bf z}(\tau_-^x)$ and $r=|{\bf r}|$. 
It is the spacial distance for the observer moving with the particle.

The particle's trajectory is fluctuating around the classical trajectory 
and can be expressed as
$z=z_0 + \delta z+ \delta^2 z +\cdots$ where the expansion
is given with respect to $e$.
In terms of the expansion of the spacial distance
$\rho = \rho_0 + \delta \rho + \delta^2 \rho + \cdots$,
the inhomogeneous solution (\ref{trans-inh}) becomes
\begin{align}  
 \phi_{inh} 
 = \frac{e}{4\pi \rho_0} 
 \left( 1 -\frac{\delta\rho}{\rho_0} 
   + \left( \frac{\delta\rho}{\rho_0} \right)^2 
	- \frac{\delta^2 \rho}{\rho_0} + \cdots 
 \right).
\end{align}
The first term is the classical potential, but
since the particle's trajectory  deviates from the classical one,
the potential also receives corrections.
Here $\rho_0$, $\delta \rho$ and $\delta^2 \rho$ are given by
\begin{align}
  \rho_0 =& 
  \dot{z}_0 (\tau^x_-) \cdot (x -z_0 (\tau^x_-) ) \\
  \delta \rho =& 
  \delta \dot{z} (\tau^x_-) \cdot (x -z_0 (\tau^x_-) ) 
  -\dot{z}_0(\tau^x_-) \cdot \delta z(\tau^x_-),\\
  \delta^2 \rho =& 
  \delta^2 \dot{z} (\tau^x_-) \cdot (x -z_0 (\tau^x_-) ) 
  -\delta \dot{z}(\tau^x_-) \cdot  \delta z(\tau^x_-)
  -\dot{z}_0(\tau^x_-) \cdot \delta^2 z(\tau^x_-).
\end{align}
From now on we only consider the transverse fluctuations.
Then $\dot{z}_0 \cdot \delta z=0$ is satisfied.
As seen from (\ref{rhoLorentz}), $\rho$ is proportional to
the spacial distance from the particle to the observer.
The variation of $\rho$ becomes negligible 
for large distance $r$ if we take 
a variation of $(x-z_0(\tau))$ in $\rho.$
On the contrary, if we take a variation of $\dot{z}_0$,
$\delta \rho$ or $\delta^2 \rho$ is still proportional to
the spacial distance $r$.
 Hence for large $r$, we can approximate the variations by
\begin{align}
  \delta \rho \sim  \delta \dot{z} (\tau^x_-) \cdot (x -z_0 (\tau^x_-) ), \\ 
  \delta^2 \rho \sim
  \delta^2 \dot{z} (\tau^x_-) \cdot (x -z_0 (\tau^x_-) ). 
\end{align}
Note also  $\langle \delta^2 \dot{z}^i \rangle = 0$ since
the velocity in the transverse directions fluctuates 
uniformly and its expectation value vanishes.
\subsection{Correlations of the inhomogeneous terms}
Now we calculate the 2-point function explicitly. 
If we take the classical part without the fluctuation of $\rho$, 
the 2-point function becomes
\begin{align}
 G(x,x') - G_0 (x,x') \rightarrow 
 \langle \phi_{inh}(x) \phi_{inh}(x') \rangle
 =  \left( \frac{e}{4\pi} \right)^2 \frac{1}{\rho_0(x) \rho_0(x')}.
\end{align}
This gives the classical radiation corresponding to the Larmor radiation.
The interference term vanishes because the 1-point function vanishes identically
$\langle \phi_h \rangle =0$. 

Corrections to the above  classical Larmor radiation are induced 
by the transverse fluctuating motion
 $\delta \rho$. 
First we consider the 2-point correlation function between the inhomogeneous terms
up to the second order of the transverse fluctuations.
Since $\langle \delta^2 \rho \rangle =0$, we have
\begin{align}
  \langle \phi_{inh}(x) \phi_{inh}(y) \rangle 
  = 
  \left(\frac{e}{4\pi} \right)^2 &
  \left\langle \frac{1}{\rho(x)\rho(y)} \right\rangle \nonumber \\
  =  
  \left(\frac{e}{4\pi} \right)^2 &\frac{1}{\rho_0(x)\rho_0(y)} 
  \biggl( 
    1 + 
	 \frac{\langle \delta \rho(x) \delta \rho(y)\rangle}{\rho_0(x)\rho_0(y)} + 
	 \frac{\langle ( \delta\rho(x) )^2 \rangle}{\rho_0^2(x)} + 
	 \frac{\langle ( \delta\rho(y) )^2 \rangle}{\rho_0^2(y)} 
  \biggr).
\label{inh-inh-result}
\end{align}
Note that all the terms in the parenthesis behave constantly
as the distance $r$ between the observer and the particle becomes
large.
The first term  gives the Larmor radiation mentioned above.
The other terms correspond to the radiation induced by the  fluctuations.
Its calculation  is easy, because one can write
$\langle \delta \rho \delta \rho \rangle$ in terms of 
$\langle \delta\dot{z}^i\delta\dot{z}^i \rangle =\langle \delta v^i \delta v^i\rangle$
which we have already evaluated in the previous section.
With the expression (\ref{momentum-fluctuation}), it becomes
\begin{align}
  \langle \phi_{inh}(x) \phi_{inh}(y) \rangle
  = &
  \left( \frac{e}{4\pi} \right)^2 \frac{1}{\rho_0(x)\rho_0(y)}
\biggl[ 1 +
  e^2 
  \int \frac{d\omega}{2\pi} 
  |h(\omega)|^2 I(\omega)
  \n 
  & \qquad \qquad \qquad
  \times \Bigl( 
     \frac{x^i y^i e^{-i\omega (\tau^x_- -\tau^y_-)}}{\rho_0 (x) \rho_0 (y)}
	 + \frac{x^i x^i}{\rho_0 (x) \rho_0 (x)}
	 + \frac{y^i y^i}{\rho_0 (y) \rho_0 (y)} \Bigr)  \biggr].
\label{EqIninScl}
\end{align}
As before, since we are considering the fluctuating motion whose frequency
is smaller than the acceleration, we may as well approximate 
$I(\omega)$  by $a^3/12\pi^2.$
In order to calculate the symmetrized correlation function between $x$ and $y$,
$I(\omega)$ is replaced by $I_S(\omega)$. 

 \subsection{Interference terms}
Next let us calculate the interference terms. They are rewritten as
\begin{align}
  \langle 
    \phi_{inh}(x) \phi_h(y) \rangle + \langle \phi_h(x) \phi_{inh}(y) 
  \rangle
  = 
  - \frac{e}{4\pi} 
   \left( 
	   \frac{\langle \delta\rho(x) \phi_h(y) \rangle}{\rho_0^2(x)} 
	 + \frac{\langle \phi_h(x) \delta\rho(y) \rangle}{\rho_0^2(y)} 
	\right).
\end{align}
Calculation of the
interference terms are more complicated since we need to evaluate 
the following correlation functions: 
\begin{align}
  \langle \delta \rho(x) \phi_h(y) \rangle 
  =& 
  - x^i \langle \delta \dot{z}^i(\tau^x_-) \phi_h(y) \rangle \\
  =&
  - e x^i \int \frac{d\omega}{2\pi} e^{-i\omega \tau^x_-} h(\omega) 
  \langle \partial_i \varphi(\omega) \phi_h(y) \rangle \\
  \langle \phi_h(x) \delta \rho(y) \rangle
  =& 
  - y^i \langle \phi_h(x) \delta \dot{z}^i(\tau^y_-) \rangle \\
  =&
  -e y^i \int \frac{d\omega}{2\pi} e^{-i\omega \tau^y_-} h(\omega) 
  \langle \phi_h(x) \partial_i \varphi(\omega) \rangle.
\end{align}
Since two terms 
$\langle \partial_i \varphi(\omega) \phi_h(y) \rangle$ and 
$\langle \phi_h(x) \partial_i \varphi(\omega) \rangle$
are related by
\begin{align}
  \langle \partial_i \varphi (\omega) \phi_h(y) \rangle
  =&
  ( \langle \phi_h(y) \partial_i \varphi (-\omega) \rangle )^{*},
\end{align}
it is sufficient to calculate  one of them.
From the definition of $\varphi$ in (\ref{varphi}),
the interference term 
$\langle \phi_h(x) \partial_i \varphi(\omega) \rangle$ is written as
\begin{align}
  \langle \phi_h(x) \partial_i \varphi (\omega) \rangle
  =& 
  \int d\tau e^{i\omega \tau} 
  \left(
    \frac{\partial}{\partial y^i} \langle \phi_h(x) \phi_h(y) \rangle 
  \right)_{y=z(\tau)} \\
  =& 
  - \frac{1}{4\pi^2} \int d\tau e^{i\omega \tau} 
  \left( 
    \frac{\partial}{\partial y^i} \frac{1}{(x^0-y^0-i\epsilon)^2 
	 - (\overrightarrow{x}- \overrightarrow{y})^2} 
  \right)_{y=z(\tau)} \\
  =& 
  \frac{1}{4\pi^2} \frac{\partial}{\partial x^i} 
    \int d\tau \frac{e^{i\omega \tau} }
    {(x^0 -z^0(\tau) -i\epsilon)^2 -(x^1 -z^1(\tau))^2 - x_{\perp}^2 }, 
\end{align}
where $x_\perp ^2 = (x^2)^2 +(x^3)^2$ is the transverse distance. 
We first evaluate the integral and then take a derivative. The integral
\begin{align}
  P(x,\omega) \equiv & \int d\tau \frac{e^{i\omega \tau} }
   {(x^0 -z^0(\tau) -i\epsilon)^2 -(x^1 -z^1(\tau))^2 -x_\perp^2},
\end{align}
can be evaluated by the contour integral in the complex $\tau$ plane.
The positions of the poles are given by a series of points 
\begin{align}
  \tau_\pm^n =& T_{\pm} + \frac{2n \pi i}{a}-i\epsilon,
\end{align}
where $n$ is an integer. $T_\pm$ are complex numbers
whose imaginary parts are $0$ or $\pi/a$ and satisfy
\begin{align}
  e^{aT_{\pm}} =& 
  \frac{a}{2u} \left(-L^2 \mp 
  \sqrt{L^4 + \frac{4}{a^2}uv} \right).
\end{align}
Here we have defined
\begin{align}
  L^2 =& -x^\mu x_\mu + \frac{1}{a^2},\\
  u =& x^0 - x^1, \quad v = x^0 + x^1.
\end{align} 
Note the relation $e^{aT_+} e^{aT_-} =-v/u$.
The positions of the poles reflect the finite temperature property of 
the uniformly accelerated  observer.
In the following we will consider two different types 
of observers as shown in 
Fig.\ref{FigRind}. 
The first observer  is to observe the radiation 
in the right wedge ($O_{\text{R}}$) while the second one is
 in the future wedge ($O_{\text{F}}$). 
For both cases, $v>0$ is satisfied and the radiation
can travel causally from the particle to the observers. 
There are two different types of poles $\tau_{\pm}$. 
A pole at $\tau_-=T_-$, which is real,
is located at a classically acceptable point.
Namely, $\tau_-$ is the proper time of 
the particle whose radiation travels to the 
observer in a causal way.
The other pole at $T_+$ is more subtle. 
\begin{figure}[p] 
\begin{center}
  \includegraphics{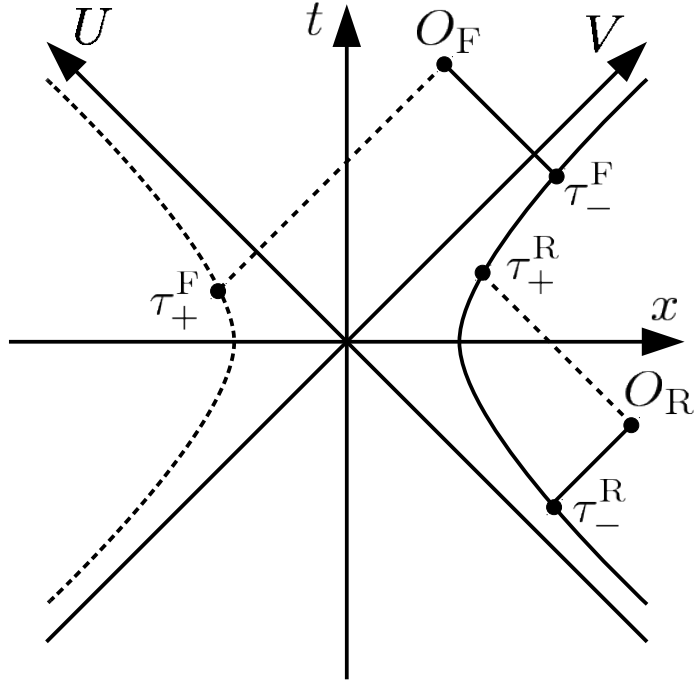}
  \caption{The hyperbolic line in the right wedge denotes the world line of the particle. 
The points $O_{\text{F}}$ and $O_{\text{R}}$ are observers in the future and right wedges, respectively.
For an observer in  the right wedge, the light-cone of the observer has two intersections with the world line, and the proper time of the intersections is
given by $\tau^R_{\pm}$. 
For an observer in  the future wedge, there is only one intersection 
on the particle's real trajectory which corresponds to 
$\tau^F_-$.
The other solution $T_+^{\text{F}}=\tau_+^{\text{F}}+i \pi/a $ is complex. 
One may interpret this complex proper time  as
 the intersection between the light-cone of the observer and the world line 
of a virtual particle with a real proper time $\tau_+^{\text{F}} $ 
in the left wedge. The superscript letters $R$ or $F$ are used to distinguish
 two different observers, but we do not use them in the body of the paper
to leave the space for the observer's position $x$. 
 }
  \label{FigRind}
\end{center}
\end{figure} 
For $u < 0$ (in the right wedge), $T_+=\tau_+^R$ is real and  corresponds to 
the advanced causal proper time. 
For $u > 0$ (in the future wedge), 
$T_+=\tau_+^F + i \pi/a$ has an imaginary part 
and one can interpret it as the proper time of a trajectory of 
a virtual particle in the left wedge, 
as in Fig. \ref{FigRind}. 
In the following, we drop the superscript $F$ or $R$.
In the region where $v<0$, $\phi_{inh}$ does not exist
and  no nontrivial correlation is observed there.

The residue of the pole at $\tau_\pm^n$ is given by 
$-e^{i \omega \tau_\pm^n}/(2\rho(\tau_\pm^n))$ where
\begin{align}
\rho(\tau_\pm^n) 
&= \dot{z}(\tau_\pm^n) \cdot (x -z_0(\tau_\pm^n)) \\
&= \frac{1}{2}(u e^{a \tau_\pm^n} +v e^{-a \tau_\pm^n}).
 \end{align}
Because of the periodicity, $\rho(\tau_\pm^n)$ is independent of $n$.
The integral is now given by
\begin{align}
  P(x,\omega) =&
    \frac{-\pi i}{\rho_0} \frac{1}{e^{2\pi \omega/a} -1}
	 \bigl( 
	   e^{i\omega \tau_-^x} 
		-e^{i\omega \tau_+^x} Z_x(\omega)
	 \bigr),
	  \label{integralP}
\end{align}
where
\begin{align}
Z_x(\omega) = e^{\pi \omega /a} \theta (u) + \theta (-u) 
\end{align} 
 $\rho_0 = \rho(\tau_-^n)$ can be rewritten in terms of $L^2$ as
\begin{align}
  \rho_0   =  \frac{a}{2}\sqrt{ L^4 +\frac{4}{a^2} uv } .
\end{align}
Note that the relation $\rho(\tau_+^n) =-\rho_0$ 
follows the identity $e^{aT_+} e^{aT_-} =-v/u$.
 
The second term of the parenthesis in (\ref{integralP}) depends on 
$\tau^+$. With naive intuition based on  classical causality,
the term may be removed by hand, but the calculation of the interference
terms is essentially quantum mechanical, and it should not be neglected.
It is puzzling how we can physically interpret such $\tau^+$ dependence of 
the integral. 

Taking a derivation of $P(x, \omega)$, we obtain 
$\langle \phi_h(x) \partial_i \varphi (\omega) \rangle$
as 
\begin{align}
  \langle \phi_h(x) \partial_i \varphi (\omega) \rangle
  =&
  \frac{i a x^i}{4 \pi \rho_0{}^2} \frac{1}{e^{2\pi \omega /a} -1} 
  \biggl(
     \Bigl( \frac{a L^2}{2\rho_0} +\frac{i\omega}{a} \Bigr) 
	  e^{i\omega \tau_-^x}
    +\Bigl(-\frac{a L^2}{2\rho_0} +\frac{i\omega}{a} \Bigr) 
	  e^{i\omega \tau_+^x} Z_x (\omega)
  \biggr).
\end{align}
Here we have used the following identities,
\begin{align}
  \frac{\partial \rho_0}{\partial x^i} 
  =&
  \frac{a^2 L^2}{2\rho_0} x^i ,\\
  \frac{\partial \tau^x_{\pm}}{\partial x^i}
  =&
  \pm \frac{1}{\rho_0} x^i,
\end{align} 
where $i$ is the transverse direction. 
The second identity can be obtained by differentiating 
$(x-z(\tau_\pm^x))^2=0$ with respect to $x^i$, see 
(\ref{difftau}).

The whole interference terms are now given by
\begin{align}
 &\langle \phi_h (x) \phi_{inh} (y) \rangle 
  + \langle \phi_{inh} (x) \phi_h (y) \rangle \n
& =
  \frac{ -i a e^2 x^i y^i }{ (4 \pi)^2 \rho_0(x)^2 \rho(y)^2 }
 \int \frac{d\omega}{2\pi} \frac{1}{1 -e^{-2\pi \omega/a}} \Bigl[
    e^{-i\omega (\tau_-^x -\tau_-^y)} 
	 \left(
	   h(-\omega) \left( \frac{aL_x^2}{2 \rho_0(x)} -\frac{i\omega}{a} \right)
	 - h( \omega) \left( \frac{aL_y^2}{2 \rho_0(y)} +\frac{i\omega}{a} \right) 
	 \right) 
  \n & 
	+ e^{-i\omega (\tau_+^x -\tau_-^y)} h(-\omega) 
	  \left( - \frac{a L^2_x}{2 \rho_0(x)} -\frac{i\omega}{a} \right)
	  Z_x (-\omega) 
	- e^{-i\omega (\tau_-^x -\tau_+^y)} h(\omega) 
	  \left( - \frac{a L^2_y}{2 \rho_0(y)} +\frac{i\omega}{a} \right) 
	  Z_y (-\omega)
  \Bigr].
\label{EqInterfereScl}
\end{align}
\subsection{Partial cancellation}
In the following, in order to see whether there is a cancellation
between the interference terms and the correlation function 
of the inhomogeneous terms, we look more closely at the 
 first term in the parenthesis of
 (\ref{EqInterfereScl}) which depends only  on $\tau_-$. 
Note that the correlation function 
of the inhomogeneous terms (\ref{EqIninScl})
depends only on $\tau_-$, and the $\tau_+$ depending terms in the 
interference terms cannot be canceled with 
the correlations of the inhomogeneous terms.

Note that, by using the relation
\begin{align}
  h(\omega) + h(-\omega) 
  =&
  \frac{e^2}{6\pi} (\omega^2 +a^2) |h(\omega)|^2,
\end{align}
one can show that a part of the interference terms in (\ref{EqInterfereScl})
\begin{align}
& \frac{ i a e^2 x^i y^i }{ (4 \pi)^2 \rho_0(x)^2 \rho_0(y)^2 }
  \int \frac{d\omega}{2\pi} \frac{1}{1 -e^{-2\pi \omega/a}} 
    e^{-i\omega (\tau_-^x -\tau_-^y)} 
	 \left(
	    h(-\omega) \frac{i\omega}{a} + h( \omega) \frac{i\omega}{a} 
    \right) 
\end{align}
can be rewritten as
\begin{align}
& -\left( \frac{e}{4\pi} \right)^2 
   \frac{ x^i y^i }{ \rho_0(x)^2 \rho_0(y)^2 }
   \int \frac{d\omega}{2\pi} e^{-i\omega (\tau_-^x -\tau_-^y)}
	 |h(\omega)|^2 I(\omega) = -\left( \frac{e}{4\pi} \right)^2 
    \frac{\langle \delta \rho(x) \delta \rho(y) \rangle}{\rho_0(x) \rho_0(y)}.
\end{align}
This  cancels the first correction term in the correlations of the inhomogeneous parts in 
(\ref{EqIninScl}). Note that this canceled term
was obtained by taking a derivative of $e^{i \omega \tau_-^x}$ in $P(x,\omega)$.

We have seen that there is a partial cancellation between the interference term
and the correlations of the inhomogeneous terms, but the other terms 
are not canceled each other.
Then, summing up both  contributions, (\ref{EqIninScl}) and 
 (\ref{EqInterfereScl}), we get 
 the following result of  the 2-point function;
\begin{align}
   \langle \phi(x) \phi(y) \rangle - \langle \phi_h(x) \phi_h(y) \rangle 
  = 
  \frac{e^2}{(4\pi)^2\rho_0(x)\rho_0(y)} F(x,y)
  \label{twopff}
\end{align}  
  where
\begin{align}
 & F(x,y) =
    1 + e^2 \int \frac{d\omega}{2\pi} \frac{|h(\omega)|^2}{6\pi} 
 I(\omega)
    \biggl(
	   \left( \frac{x^i}{\rho_0 (x)} \right)^2
	 + \left( \frac{y^i}{\rho_0 (y)} \right)^2
	 \biggr) \n &
- \frac{ia^2 x^i y^i}{\rho_0(x)\rho_0(y)}  
  \int \frac{d\omega}{4\pi} \frac{1}{1 -e^{-2\pi \omega/a}} 
  \biggl[
    e^{-i\omega (\tau_-^x -\tau_-^y)} 
	 \left(
	   h(-\omega) \frac{L_x^2}{ \rho_0(x)} 
	  -h( \omega) \frac{L_y^2}{ \rho_0(y)} 
	 \right)
     \n & 
  -e^{-i\omega (\tau_+^x -\tau_-^y)} h(-\omega) 
  \left( 
    \frac{ L^2_x}{ \rho_0(x)} +i \frac{2\omega}{a^2} 
  \right) Z_x (-\omega)
  -e^{-i\omega (\tau_-^x -\tau_+^y)} h(\omega) 
  \left( 
    - \frac{ L^2_y}{ \rho_0(y)} +i \frac{2\omega}{a^2} 
  \right) Z_y (-\omega )
  \biggr]
\label{twopff2}.
\end{align}
The first term in $F(x,y)$ is the classical effect of radiation
corresponding to the Larmor radiation.
The second term comes from the inhomogeneous term
$ \langle (\delta \rho(x)/\rho_0(x))^2  \rangle
+ \langle (\delta \rho(y)/\rho_0(y))^2 \rangle$.
The third term comes from the interference term,
which is obtained by taking a derivative of 
$\rho(x)$ in $P(x,\omega)$.
The fourth term is also an interference effect and depends on $\tau_+$.

Let us compare the above result with the calculation for an internal detector.
In the case of an internal detector in (1+1) dimensions,
the radiation  
is canceled by the interference effect, and
there are no terms depending on $\tau_+$. 
In the case of an internal detector in (3+1) dimensions,
there are $\tau_+$-dependent terms.
But if we neglect these terms,
it was shown \cite{JohnsonHu} that 
the interference terms completely cancel the radiation.
The calculation is reviewed in Appendix A.
In the case of a charged particle,
since the position of the particle is fluctuating,
only a part of the terms is canceled. 
In the following we focus on the $\tau_+$-independent terms
because the rapidly oscillating function $e^{-i\omega \tau}$ 
remains even after setting $x=y$ and
suppresses the  $\omega$ integral of the $\tau_+$-dependent terms.

For a symmetrized 2-point function, $F(x,y)$
is replaced by $F_S(x,y)$
\begin{align}
 F_S(x,y) =&
    1 + e^2 \int \frac{d\omega}{2\pi} \frac{|h(\omega)|^2}{6\pi} 
 I_S(\omega)
    \biggl(
	   \left( \frac{x^i}{\rho_0 (x)} \right)^2
	 + \left( \frac{y^i}{\rho_0 (y)} \right)^2
	 \biggr) \n &
- \frac{ia^2 x^i y^i}{\rho_0(x)\rho_0(y)}  
  \int \frac{d\omega}{8\pi} 
  \coth \left(\frac{\pi \omega}{a} \right)  
    e^{-i\omega (\tau_-^x -\tau_-^y)} 
	 \left(
	   h(-\omega) \frac{L_x^2}{ \rho_0(x)} 
	  -h( \omega) \frac{L_y^2}{ \rho_0(y)} 
	 \right)
   \n & 
	+ \text{$\tau_+$-dependent terms}
\label{twopff3}.
\end{align}  

\subsection{Energy-momentum tensor}
In the remainder of this section
 we consider the radiation emitted by the accelerated particle. 
The energy-momentum tensor of the scalar field is given by
\begin{align}
  \langle T_{\mu\nu} \rangle 
  = 
 \langle :\partial_{\mu}\phi \partial_{\nu}\phi 
  -\frac{1}{2} g_{\mu\nu} \partial^{\alpha}\phi \partial_{\alpha}\phi: \rangle_S.
\end{align}
Hence we can evaluate it  by taking a derivative of 
the 2-point function (\ref{twopff3}).

The following relations are useful in taking derivatives:
\begin{align}
  \partial_\mu \rho_0 
  =& (\ddot{z_0} \cdot (x-z_0) -1) \partial_\mu \tau_- + \dot{z}_{0\mu} \\
  =& -\frac{a^2 L^2}{2} 
   \, \partial_\mu \tau_- + \dot{z}_{0\mu} 
    \\
   \partial_\mu \tau_- =& \frac{x_\mu - z_{0\mu}}{\rho_0}  
 \label{dtaudrho}
\end{align}
In the last line of the first equation, 
we used the explicit form of the classical solution
 (\ref{classical-trajectory}) and $\ddot{z}_0 \cdot x = - a^2L^2/2$. 
The derivative $\partial_\mu \tau_-$ was
obtained  by
taking a variation of the light-cone condition $(x-z_0(\tau^x_-))^2=0$;
\begin{align}
  2 (x_\nu - z_{0\nu}) (\delta x^\nu - \dot{z}^\nu_0 \delta \tau^x_-) = 0 
  \longrightarrow \frac{\delta \tau^x_-}{\delta x^\nu} 
  = 
  \frac{x_\nu - z_{0\nu}}{\rho_0}.
\label{difftau}
\end{align}
In particular, $u$ and $v$ derivatives are given by
\begin{align}
 \partial_u \tau_- &=\frac{v-v_z}{2\rho_0}, &
 \partial_v \tau_- &=\frac{u-u_z}{2\rho_0} \\
 \partial_u \rho_0 &= -\frac{a^2 L^2}{2} \partial_u \tau_- + \frac{a v_z}{2},& 
\partial_v \rho_0 &= -\frac{a^2 L^2}{2} \partial_v \tau_- + \frac{1}{2a v_z}
\end{align}
where
$u_z=-e^{-a \tau_-}/a, v_z=e^{a \tau_-}/a$.
From (\ref{dtaudrho}), we have 
$(\partial \rho_0)^2 =  a^2 x^2$. Since $(x-z(\tau))^2=0$,
$x^2 \sim {\cal O}(r)$ and $(\partial \rho_0)^2$ is approximately
proportional to the spacial distance $r$, not $r^2$.
On the other hand, since $L^2=-x_\mu^2+1/a^2$ is ${\cal O}(r)$, 
$\partial_\mu \rho_0$ itself is growing as ${\cal O}(r)$.

First we calculate the classical part of the energy-momentum tensor.
It becomes
\begin{align}
  T_{\text{cl}, \mu\nu} 
  =&
  \frac{e^2(\partial_\mu \rho_0 \partial_\nu \rho_0 -\frac{g_{\mu\nu}}{2} \partial_\alpha \rho_0 \partial^\alpha \rho_0)}
  {(4\pi)^2 \rho^4_0}  \\
 \sim &
  \frac{e^2 \partial_\mu \rho_0 \partial_\nu \rho_0 }
  {(4\pi)^2 \rho^4_0}.
\end{align}
Note that $\partial_\alpha \rho_0 \partial^\alpha \rho_0$ does not 
make a contribution here,
since it is of the order of $\rho_0$ at infinity while $\partial_\mu \rho_0 \partial_\nu \rho_0$ 
is in general of order $\rho^2_0$.
This part of the energy-momentum tensor corresponds to the classical 
Larmor radiation and behaves as $1/\rho_0^2 \sim 1/r^2$ at infinity.
The term $\dot{z}_{0\mu}(\tau^x_-)$ in $\partial_\mu \rho_0$ seems 
to be negligible, since it is ${\cal O}(1)$ while 
$\partial_\mu \rho_0$ is ${\cal O}(r)$. 
However, care should be taken because 
 $\dot{z}_{0\mu}(\tau^x_-)= (\cosh a\tau^x_-, \sinh a \tau^x_-,0,0)$ 
behaves singularly if the observer is  near the horizon.

Next we evaluate the other parts of the energy-momentum tensor.
We especially consider the $(u,u)$ and $(v,v)$-components in the following. 
From (\ref{twopff3}), extra terms of the energy-momentum tensor 
besides the classical ones are given by
\begin{align}
  T_{\text{fluc},\mu\nu} 
  = &
  \frac{ (x^i)^2 }{ \rho^2_0 } 
  \biggl[
    \Bigl(
      \frac{e^2}{\pi} I_m -  \frac{6ma^2 I_1 L^2}{\rho_0} 
	 \Bigr) 
    T_{\text{cl},\mu\nu}  
	 -\frac{e^2 a^2 L^2}{(4\pi)^2 \rho_0^3}
	 \Bigl(
	   m I_3 \ \partial_{\mu} \tau_-^x \partial_{\nu} \tau_-^x 
 \n & \qquad \qquad 
		 + \frac{2m I_1}{\rho_0 L^2} 
		  ( x_{\mu} \partial_{\nu} \rho_0 +x_{\nu} \partial_{\mu} \rho_0 ) 
		  +\frac{e^2 I_m}{12\pi L^2} 
		  ( x_{\mu} \partial_{\nu} \tau_-^x +x_{\nu} \partial_{\mu} \tau_-^x )
 \n & \qquad \qquad
	  -\frac{e^2 I_m}{24\pi \rho_0} 
	   ( \partial_{\mu} \tau_-^x \partial_{\nu}\rho_0 
	   + \partial_{\nu} \tau_-^x \partial_{\mu}\rho_0 )
	 \Bigr)
  \biggr]
  \label{Tfluc}
\end{align}
where we have defined the following $\omega$ integrals
\begin{align}
 I_1 &=
   \int \frac{d\omega}{4 \pi} |h(\omega)|^2 
	\coth \left(\frac{\pi \omega}{a} \right) \ \omega , \\
 I_3 &=
   \int \frac{d\omega}{4 \pi} |h(\omega)|^2 
	\coth \left(\frac{\pi \omega}{a} \right) \ \omega^3 ,\\
 I_m &=
  \int \frac{d\omega}{4 \pi} |h(\omega)|^2 
  \coth \left(\frac{\pi \omega}{a} \right) \ 
  (\omega^3 + a^2 \omega) \\
  &=I_3+ a^2 I_1 .  
\end{align}
These integrals can be similarly evaluated as in Sec.~III, and we have
\begin{align}
I_1 & = \frac{3}{2mae^2} ,\\ 
I_m &\sim a^2 I_1 .
\end{align}
Because of the inequality $\Omega_- \ll a$, terms containing 
$I_3 $ are
generally negligible compared to other terms; 
$I_3 \sim \Omega_-^2 I_1 \ll a^2 I_1.$

Near the past horizon, the $v\rightarrow 0$, the $u$-derivatives
of $\rho_0$ and $\tau^x_-$ become very small and negligible.
On the other hand, $v$-derivative of $\tau_-$ becomes potentially large.
$u$-derivatives of them are approximately given by
\begin{align}
  \partial_v \tau^x_- \rightarrow \frac{1}{av} 
  ,  \qquad
    \partial_v \rho_0 \rightarrow -\frac{au}{2}.
\end{align}
A singular term of $\partial_v \rho_0$ near $v \sim 0$ is canceled and 
it remains finite near the past horizon.
Hence the second term in (\ref{Tfluc}) proportional to 
$(\partial \tau_-)^2$ may becomes large there.
However, there are two reasons that the term cannot grow so large.
One is a suppression by the $\omega$ integral, which is proportional to a
very small coefficient $I_3$.
The other reason is the overall factor $(x^i)^2 /\rho_0^2$.
Since the observer is much further than the acceleration scale $1/a$
from the particle, $L^2$ is much larger than $1/a^2$.
Then $\rho_0=(a/2) \sqrt{L^4 +(4/a^2) uv}$ can be approximated by
$\rho_0 \sim (a/2) |x_\mu|^2$ and $(x^i)^2/\rho_0^2$ is also
suppressed. 
Because of these two reasons, the singular behavior near the past horizon
seems to be difficult to be observed experimentally. 
\section{Thermalization in Electromagnetic Field} 
\setcounter{equation}{0}
In this section, we consider the thermalization of an accelerated charged 
particle in the  electromagnetic field. 
Calculations of the energy-momentum tensors are more involved
and   left for a future investigation. 
We study the thermalization of the transverse momenta 
of a uniformly accelerated particle in an electromagnetic field.
The calculation is almost the same, but due to the presence
of the polarization, several quantities become 
twice as large as those in the scalar case. 

The action is given by 
\begin{align}
 S_{EM} = - m \int d\tau \sqrt{\dot{z}^\mu \dot{z}_\mu}
  - \int d^4 x \ j^\mu (x) A_\mu(x)
  - \frac{1}{4} \int d^4 x \ F^{\mu\nu}F_{\mu\nu},
\end{align}
where the current is defined as
\begin{align}
 j^\mu (x) = e\int d\tau
   \dot{z}^\mu(\tau)\, \delta^4(x-z(\tau)) .
\end{align}
The equations of motion are
\begin{align}
 m\ddot{z}_\mu =& eF_{\mu\nu} \dot{z}^\nu \\
 \partial_\mu F^{\mu\nu}(x) =& j^\nu.
\end{align}
Using the gauge
\begin{align}
 \partial^\mu A_\mu = 0,
\end{align}
the equation of motion for $A_\mu$ becomes
\begin{align}
 \partial^\mu \partial_\mu A^\nu = j^\nu.
\end{align}
One can solve this equation as
\begin{align}
 A_\mu =& A_{h\mu} + \int d^4 y \ G_R(x,y) j_\mu (y) \n
 =& A_{h\mu} + e\int d\tau G_R(x,z(\tau)) 
                \dot{z}_\mu(\tau),
\end{align}
where $A_{h\mu}$ is the homogeneous solution of the equation of motion which
satisfies 
$\partial^2 A_h^\mu = 0$. $G_R(x-y)$ is the retarded Green function
\begin{align}
 G_R(x,y) = \theta(x^0-y^0) \frac{\delta((x-y)^2)}{2\pi}, \qquad
  \partial^2 G_R(x,y) = \delta^4(x-y).
\end{align}
Inserting the solution of $A_\mu(x)$ back to the equation of motion for
$z^\mu$, we obtain the following stochastic equation
\begin{align}
 m\ddot{z}_\mu(\tau) =& F_\mu +
  e( \partial_\mu A_{h\nu}(z) - \partial_\nu A_{h\mu}(z) ) \dot{z}^\nu
  \n
  &+ e^2 \int d\tau^\prime \dot{z}^\nu(\tau)
  ( \dot{z}_\nu(\tau^\prime) \partial_\mu
  - \dot{z}_\mu(\tau^\prime) \partial_\nu )
  G_R(z(\tau),z(\tau^\prime)).
\end{align}
The second line is the radiation reaction which can be treated 
similarly to the
scalar case. It becomes
\begin{align}
 & e^2 \int d\tau' \
  \left( \dot{z}^\nu \dot{z}_{[\nu} \partial_{\mu]}\
  - \frac{s^2}{2}
  (\ddot{z}^2\partial_\mu - \dddot{z}_\mu \dot{z}^\nu \partial_\nu) \right)
  \frac{d}{ds} G_R(z(\tau),z(\tau^\prime)) \nonumber \\
 =& - e^2 \int^\infty_\infty ds \ \frac{s^2}{3}
  ( \dddot{z}_\mu(\tau) + \dot{z}_\mu(\tau) \ddot{z}^2(\tau) )
  \frac{d}{ds} \frac{\delta(s^2)}{2\pi} \nonumber \\
 =& \frac{e^2}{6\pi} ( \dddot{z}_\mu + \dot{z}_\mu \ddot{z}^2 ),
\end{align}
which has exactly the same form as the scalar case,
but the coefficient is twice as large 
since the gauge field has two different polarizations. 
This is the  Abraham-Lorentz-Dirac  self-radiation term.

For small transverse momentum fluctuations 
$ \delta v^i \equiv \delta \dot{z}^i$,
we can simplify the stochastic equation similarly to the scalar case in previous sections.
It can be solved
in terms of  the homogeneous solution of the gauge field as 
\begin{align}
 \delta\tilde{v}^i(\omega) = -eh(\omega)
  ( v_{0\alpha} \partial_i
  + \delta^i_\alpha (v_0 \cdot \partial) )
  A^\alpha_h, 
  \label{EMdeltav}
\end{align}
where 
\begin{align}
 h(\omega) = \frac{1}{-im\omega + \frac{e^2}{6\pi}(\omega^2 + a^2)}.
\end{align}
The noise correlation of $ A_h^\mu$
in the r.h.s. of (\ref{EMdeltav})  can be evaluated as 
\begin{align}
  (v_{0\alpha} \partial_i
  + \delta^i_\alpha (v_0 \cdot \partial) )
  (v'_{0\beta} \partial'_j
  + \delta^j_\beta (v'_0 \cdot \partial'))
  \langle A^\alpha_h(z)A^\beta_h(z^\prime) \rangle =
  \frac{a^4}{16\pi^2} 
  \frac{\delta_{ij}}{\sinh^4 \left(\frac{a(\tau-\tau^\prime-i\epsilon)}{2}\right)}.
\end{align}
It is also twice as large as the scalar case. 
Note that the quantity is gauge invariant
\begin{align}
 (\dot{z}_\alpha k_\mu
  -\eta_{\alpha \mu} (\dot{z} \cdot k))
 (\dot{z}^\prime_\beta k^\prime_\nu
  -\eta_{\beta \nu} (\dot{z}' \cdot k')) k^\alpha k^{\prime \beta} = 0.
\end{align}
Hence performing similar calculations to the scalar case,
the fluctuations of the transverse momenta become
\begin{align}
 \frac{m}{2} \langle \delta v^i(\tau) \delta v^j(\tau) \rangle
  = \frac{1}{2}\frac{a \hbar}{2\pi c} \delta_{ij} 
    \left( 1 +\mathcal{O}\left( \frac{a^2}{m^2} \right) \right).
\end{align}
Since the coefficient of the dissipative term is twice as large
as the scalar case, 
the relaxation time becomes the half of it:
$\tau_R=\frac{6\pi m}{a^2 e^2}$.
\section{Conclusions and Discussions} 
\setcounter{equation}{0}
In this paper, we studied a stochastic motion of a
uniformly accelerated charged particle in the scalar-field analog of QED.
The particle's motion fluctuates because of the thermal behavior
of the uniformly accelerated observer (the Unruh effect).
Because of this fluctuating motion, Chen and Tajima
\cite{ChenTajima}
conjectured that there is additional 
radiation besides the classical Larmor radiation. 
On the other hand, it was  argued \cite{Sciama,RavalHuAnglin}
that interferences between the radiation field induced by
the fluctuating motion and the quantum fluctuation of the vacuum
may cancel the above additional radiation. 
The cancellation was shown in the case of an 
internal detector, but it was not yet settled
whether the same kind of cancellation occurs in the 
case of a fluctuating charged particle in QED.

In the present paper, in order to investigate the above issue 
systematically, we first formulated a motion of a uniformly accelerated
particle in terms of the stochastic (Langevin) equation. By using this formalism,
we showed that the momenta in the transverse
directions actually get thermalized so as to satisfy the equipartition
relation with the Unruh temperature.
Then we calculated correlation functions and energy flux from the 
accelerated particle. Partial cancellation is actually 
shown to occur,  
but some terms still remain. Hence 
there is still a possibility that, besides the
classical Larmor radiation, we can detect additional
radiation associated with the fluctuating motion 
caused by the Unruh effect. 

There are several issues to be clarified.
First in calculating the energy flux at infinity
there appeared classically unacceptable contributions
(i.e. those depend on $\tau_+$).
If the observer is in the right wedge, the contribution
to the energy flux come from the particle in the future
of the observer. In the case of the observer in the future
wedge, this contribution comes from the virtual particle
in the left wedge. Both of them are classically unacceptable,
and we do not yet have  physical understanding 
why these contributions appear in the calculation. 

Another issue is the calculation of longitudinal 
fluctuations. Since the particle is accelerated in the longitudinal direction
with very high acceleration,
longitudinal fluctuations caused by the Unruh effect are technically difficult to evaluate
(see Appendix \ref{app2}). 
Furthermore, it is not clear what kind of quantities are thermalized in the 
longitudinal fluctuations.
The particle feels finite temperature noise in the accelerating frame
and we may expect that the kinetic energy written in the Rindler coordinate $\xi$
is thermalized.
However, the particle is uniformly accelerated in  constant external force.
Then if a particle at $P_1$ on a trajectory $T_1$ in Fig.~\ref{yamamoto}
is kicked in the longitudinal direction to  $P_2$,
it follows another trajectory $T_2$ with a different asymptote.
The Rindler coordinate  $\xi$, which is 
defined in the original accelerating frame,
becomes divergent at $\tau=\infty$ on the new trajectory $T_2$.
In this sense, the longitudinal fluctuations look kinematically unstable
in the original Rindler coordinate and 
it is inappropriate to describe the thermalization of longitudinal fluctuations
in the Rindler coordinate $\xi$. 
We give  brief discussions on the 
calculations of longitudinal fluctuations in Appendix \ref{app2}.

\begin{figure}[htb] 
\begin{center}
  \includegraphics[width=20em]{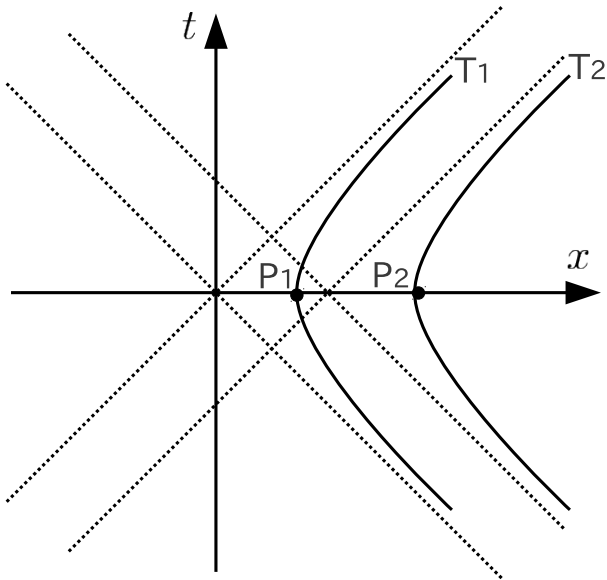}
  \caption{
$T_1$ is a trajectory of the original accelerated particle.
If the particle is kicked at $P_1$ to a point $P_2$, it follows a different
trajectory $T_2$ in the constant external force.
It has a different asymptote from the original one, and the 
original Rindler coordinate $\xi$ of the particle on $T_2$ diverges
at $\tau=\infty$.  
 }
  \label{yamamoto}
\end{center}
\end{figure} 

Finally 
an interesting possibility is an effect of decoherence
induced by interaction with environments.
In this paper, we have treated the trajectory of the particle semiclassically 
in terms of the stochastic Langevin equation.
If the initial state of the particle is a superposition of  two localized 
wave-packets, we need to sum 
over the corresponding trajectories to obtain the transition amplitude \cite{JohnsonHu}.
Decoherence would suppress the correlations between these different worldlines.
Furthermore the effect of decoherence  might be important even 
for a single trajectory with the stochastic fluctuations.
We have discussed the partial cancellation in the energy-momentum tensor 
between the inhomogeneous ( fluctuation originated) terms
and the interference terms. 
The fluctuating motions of the particle
are quantum mechanically induced by the vacuum fluctuations of the radiation field.
The inhomogeneous terms in energy-momentum tensor are evaluated by calculating the 
correlation of the vacuum fluctuation at the almost same
positions on the trajectory, and  robust against decoherence.
Namely the variances of the trajectory never vanish. 
On the contrary, the interference terms are given by calculating 
the correlation function of the vacuum fluctuations at the  position of the observer
$\cal{O}_{F,R}$ and at the particle's position on the trajectory,
and can be easily affected by an additional interaction
in between the trajectory and the observer.
Then the (partial) cancellation is lost
and  only the correlation function
between the inhomogeneous terms
$\langle \phi_{inh}(x) \phi_{inh}(y)\rangle$ 
(namely, the Unruh radiation) may survive.
\section*{Acknowledgments} 
The work was started by discussions on physics of the high-intensity lasers 
with experimentalists K. Fujii, K. Homma, T. Saeki,
T. Takahashi, T. Tauchi and J. Urakawa.
We would like to thank them  for introducing us
to this subject. 
We also acknowledge K. Itakura, Y. Kitazawa, H. Kodama 
and W.G. Unruh for useful discussions. 
A part of the work was presented at a workshop at LMU on 
November 24, 2009. We thank T. Tajima and D. Habs 
for organizing the workshop and the warm hospitality there.
The research by Y.Y. and S.Z. is
supported in part by 
the Japan Society for the Promotion of Science Research Fellowship for Young Scientists.
The research by S.I. is supported in part
by the Grant-in-Aid for Scientific Research (19540316) 
from the Ministry
of Education, Culture, Sports, Science and Technology, Japan. 
We are also supported in part by the Center for 
the Promotion of Integrated Sciences (CPIS) of Sokendai.
\appendix 
\section{Internal Detector in $3+1$ dimensions} 
In this appendix, we give a brief review of an internal detector
in (3+1) dimensions to see how the  Unruh radiation is canceled
by the interference effects \cite{HuLin}. 

The action for an accelerated internal detector
coupling with a massless scalar field is given by
\begin{align}
  S = &
    \int \, d\tau \frac{m}{2} 
    \bigl( ( \partial_{\tau}Q(\tau) )^2 -\Omega_0^2 Q^2 \bigr)
  + \int d^4 x \, \frac{1}{2} (\partial_{\mu} \phi) (\partial^{\mu} \phi) \n &
  + \lambda \int d^4 x d\tau \, Q(\tau) \phi(x) \delta^4 (x - z(\tau)),
\end{align}
where $\partial_{\tau}$ is used to denote a derivative with respect to 
the proper time $\tau$. The equations of motion are given by
\begin{align}
  & \partial^2 \phi(x) = 
    \lambda \int d\tau \, Q(\tau) \delta^4 (x - z(\tau)) \\
  &( \partial_{\tau}^2 + \Omega_0^2 ) Q(\tau) =
    \frac{\lambda}{m} \phi (z(\tau)).
\end{align}
Substituting the  solution $ \phi = \phi_h + \phi_{inh} $,
\begin{align}
  \phi_{inh} (x) = 
    \lambda \int d\tau \, Q(\tau) G_R (x-z(\tau)),
\end{align}
to the equation of the internal detector, we get the following equation,
\begin{align}
    ( \partial_{\tau}^2 + \Omega_0^2 ) Q(\tau) 
  - \frac{\lambda^2}{m} \int d\tau' \, Q(\tau') G_R (z(\tau)-z(\tau')) &=
    \frac{\lambda}{m} \phi_h (z(\tau)).
\end{align}
Here $\phi_h$ is the homogeneous solution representing 
the vacuum fluctuations.
The inhomogeneous term is evaluated by expanding the
Green function with respect to ($\tau -\tau'$) 
as we did in (\ref{EqRetexp}).
Then after a renormalization of the mass term, we get the 
diffusive term of the radiation reaction,
\begin{align}
  \int d\tau' Q(\tau') G_R (z(\tau) -z(\tau')) \ \ \Rightarrow & \ \ 
  \frac{Q'(\tau)}{4\pi}.
\end{align}
The stochastic equation can be solved 
by the Fourier transformation on the path as
\begin{align}
  \tilde{Q}(\tau) =\lambda h(\omega) \varphi (\omega), 
\end{align}
where
\be
  h(\omega)^{-1} = 
  -m\omega^2 +m\Omega^2  -  i \frac{\omega \lambda^2}{4 \pi} 
\ee
and the Fourier transformations are defined as
\begin{align}
  \tilde{Q}(\omega) =& 
    \int d\tau \, e^{i\omega \tau} Q(\tau), \\
  \varphi(\omega) =&
    \int d\tau \, e^{i\omega \tau} \phi_h (z(\tau)).
\end{align}
Note that  $G_R (z(\tau) -z(\tau'))$ is a function of 
$(\tau -\tau')$ if the classical solution 
$z(\tau)$ represents the accelerated path (\ref{classical-trajectory}).
The 2-point correlation function is decomposed into
\begin{align}
 \langle \phi(x) \phi(y) \rangle =&
   \langle \phi_h(x) \phi_h(y) \rangle +
   \langle \phi_{inh}(x) \phi_h(y) \rangle +
   \langle \phi_h(x) \phi_{inh}(y) \rangle +
\langle \phi_{inh}(x) \phi_{inh}(y) \rangle 
\end{align}
where
\begin{align}
&    \langle \phi_{inh}(x) \phi_h(y) \rangle +
   \langle \phi_h(x) \phi_{inh}(y) \rangle 
 \n 
& = \int d\tau \frac{d\omega}{2\pi} e^{-i\omega \tau} \lambda^2 h(\omega) 
 \bigl(
   G_R (y -z(\tau)) \langle \phi_h(x) \varphi(\omega) \rangle
 + G_R (x -z(\tau)) \langle \varphi(\omega) \phi_h(y) \rangle
 \bigr)  \label{app-inh-h}
\\
& \langle \phi_{inh}(x) \phi_{inh}(y) \rangle \n 
& = \int d\tau d\tau' \frac{d\omega}{2\pi} \frac{d\omega'}{2\pi} 
   e^{-i(\omega \tau +\omega' \tau')} \lambda^4 
	G_R(x -z(\tau)) G_R(y -z(\tau')) h(\omega) h(\omega')
 \langle \varphi(\omega) \varphi(\omega') \rangle .
\end{align}
We first evaluate the interference term (\ref{app-inh-h});
\begin{align}
 \langle \phi_h (x) \varphi(\omega) \rangle =&
   \int d\tau e^{i\omega \tau} \langle \phi_0(x) \phi_0(z(\tau)) \rangle \n =&
 - \frac{1}{4\pi^2} \int d\tau 
   \frac{e^{i\omega \tau} }
	     {(x^0 -z^0(\tau) -i\epsilon)^2 -(x^1 -z^1(\tau))^2 -\rho^2 }\n =&
 - \frac{1}{4\pi^2} P(x,\omega).
\end{align}
Poles are given by solving the equation,
\begin{align}
  0 =&
    \left(x^0 -\frac{\sinh a\tau}{a} \right)^2 
  - \left(x^1 -\frac{\cosh a\tau}{a} \right)^2 -\rho^2 \\ =&
  - u \frac{e^{a\tau}}{a} + u \frac{e^{-a\tau}}{a} 
  + x^2 -\frac{1}{a^2}.
\end{align}

The solutions of this equation are classified 
according to  two different types of observers (see Fig.\ref{FigRind})
\begin{align}
  \mathcal{O}_F \ (\text{in future wedge})&:\ u >0 ,\ v >0 \n 
  \Rightarrow \quad &
  e^{a\tau_-^F} = 
    \frac{a}{2u}
    \Bigl( -L^2 +\sqrt{ L^4 +\frac{4}{a^2} uv } \Bigr) \\ 
  -&e^{a\tau_+^F} = 
    \frac{a}{2u}
    \Bigl( -L^2 -\sqrt{ L^4 +\frac{4}{a^2} uv } \Bigr) \\ 
  {\cal O}_R \ (\text{in right  wedge}) &:\  u <0 ,\ x^0 +x^1 >0\n 
  \Rightarrow \quad &
  e^{a\tau_-^R} = 
    \frac{a}{2|u|}
    \Bigl( L^2 -\sqrt{ L^4 -\frac{4}{a^2}|uv| } \Bigr) \\ 
  &e^{a\tau_+^R} = 
    \frac{a}{2|u|}
    \Bigl( L^2 +\sqrt{ L^4 -\frac{4}{a^2}|uv| } \Bigr),
\end{align}
where,
$L^2 = -x^2 +1/a^2.$
The poles at $\tau_-^{F,R}$  correspond to the 
proper times at the intersections of the particle's world line 
and the past light-cone of the observer's position.
Hence they are the physically acceptable poles.
On the other hand, $\tau_+^F$ correspond to the proper time
at a point on a "virtual path" 
in the left wedge.  $\tau_+^R$ lies at an intersection
of the world line and the future light-cone of the observer. 
Both of them are classically unacceptable. 

Summing these contributions to the integral, we obtain
\begin{align}
  P(x,\omega) =&
    \frac{-\pi i}{\rho_0} \frac{1}{e^{2\pi \omega/a}-1}
	 \bigl( 
	   e^{i\omega \tau_-^x} -e^{i\omega \tau_+^x} Z_x (\omega)  
	 \bigr),
\end{align}
where
\begin{align}
  Z_x = & e^{\pi \omega /a} \theta (u) + \theta (-u) , \\ 
  \rho_0 =& \frac{a}{2} \sqrt{ L^4 + \frac{4}{a^2}uv }.
\end{align}
Using the following relation,
\begin{align}
  \int d\tau \, G_R (x -z(\tau)) f(\tau) =&
    \frac{1}{4\pi \rho_0} f(\tau_-),
\end{align}
a part of the interference term depending on 
$\tau_-^R$ or $\tau_-^F$  can be written as
\begin{align}
 \langle \phi_h (x) \phi_{inh} (y) \rangle \rightarrow \ &
   i\lambda^2 \int d\tau d\tau' \frac{d\omega}{2\pi} 
	G_R(x -z(\tau)) G_R(y -z(\tau')) e^{i\omega(\tau -\tau')}
	\frac{h(\omega)}{e^{2\pi \omega/a} -1}.
\label{01corr}
\end{align}
Similarly, we have 
\begin{align}
 \langle \phi_{inh} (x) \phi_h (y) \rangle \rightarrow \ &
   i\lambda^2 \int d\tau d\tau' \frac{d\omega}{2\pi} 
	G_R(x -z(\tau)) G_R(y -z(\tau')) e^{-i\omega(\tau -\tau')}
	\frac{h(\omega)}{1 -e^{-2\pi \omega/a}},
\label{10corr}
\end{align}
where we have used the identity
\begin{align}
 \langle \tilde{\varphi}(\omega ) \phi_h (y) \rangle =&
  \bigl( \langle \phi_h (y) \tilde{\varphi}(-\omega) \rangle \bigr)^{\ast}.
\end{align}

The correlation function of inhomogeneous terms is given by
\begin{align}
 \langle \phi_{inh}(x) \phi_{inh}(y) \rangle =&
   \lambda^4 \int d\tau d\tau' \frac{d\omega}{2\pi} \frac{d\omega'}{2\pi}
	e^{-i\omega \tau} e^{-i\omega' \tau'} 
	G_R(x -z(\tau)) G_R(y -z(\tau')) h(\omega) h(\omega') 
	\langle \tilde{\varphi}(\omega) \tilde{\varphi}(\omega') \rangle \n 
 =&
   \lambda^4 \int d\tau d\tau' \frac{d\omega}{2\pi} \frac{d\omega'}{2\pi}
	e^{-i\omega (\tau -\tau')}
	G_R(x -z(\tau)) G_R(y -z(\tau')) h(\omega) h(-\omega) \n & \times 
	\int d(\tau_a -\tau_b) 2\pi \delta(\omega +\omega')  
	e^{i\omega (\tau_a -\tau_b)}
	\langle \phi_0(z(\tau_a)) \phi_0(z(\tau_b)) \rangle \n 
 =&
   \lambda^4 \int d\tau d\tau' \frac{d\omega}{2\pi} 
	e^{-i\omega (\tau -\tau')}
	G_R(x -z(\tau)) G_R(y -z(\tau')) 
	\frac{\omega}{2\pi} \frac{h(\omega) h(-\omega)}{1 -e^{-2\pi \omega /a}}.
\label{11corr}
\end{align}
These three contributions (\ref{01corr}), (\ref{10corr}), (\ref{11corr}) 
to the correlation function are shown to be canceled each other
because of the relation
\begin{align}
  h(\omega) -h(-\omega) 
  = \frac{i \omega \lambda^2 }{2\pi} |h(\omega)|^2.
\end{align}
Therefore if we neglect the contributions from the classically
unacceptable poles at $\tau^+$ the 2-point function
vanishes, and therefore there are no energy-momentum flux 
after the thermalization occurs.
 
The remaining term in the 2-point function is the
 contributions of the $\tau_+$ dependent terms
to the interference term, and written as
\begin{align}
  \int \frac{d\omega}{2\pi} \frac{-i a^2\lambda^2}{8\pi \rho_0(x) \rho_0(y)}
    \frac{1}{1- e^{-2\pi \omega /a} } 
	 ( h( \omega) e^{-i\omega (\tau_-(x) -\tau_+(y))} Z_y(-\omega)
	 - h(-\omega) e^{-i\omega (\tau_+(x) -\tau_-(y))} Z_x(-\omega)).
\end{align}
It looks strange why we have such a (classically unacceptable) 
term in the final result.
\section{Longitudinal fluctuations \label{app2}} 
\setcounter{equation}{0}
In this appendix we briefly
study fluctuations along the direction of the particle's classical motion.
It is convenient to define the light-cone coordinates
\begin{equation}
z^\pm = z^t \pm z^x.
\end{equation}
The classical solution in the light-cone coordinates is given by
\begin{equation}
z^\pm_{cl} = \frac{\pm e^{\pm a \tau}}{a}.
\label{classical-Rindler}
\end{equation}
In order to describe the fluctuations around the classical solution,
we define the Rindler coordinates ($\tilde{\tau}$, $\xi$) by
\begin{equation}
z^\pm = \pm \frac{e^{\pm a(\tilde{\tau} \pm \xi)}}{a} . 
\end{equation}
The classical solution corresponds to $\tilde{\tau}(\tau)=\tau$ 
and $\xi(\tau)=0$. 

Small fluctuations around the classical solution (\ref{classical-Rindler}) 
are written as 
\begin{equation}
\delta z^\pm = (\delta \tilde{\tau} \pm \delta \xi) e^{\pm a \tau}.
\end{equation}
Writing the velocities as $\delta \dot{z}^\pm = \pm v^\pm e^{\pm a\tau}$,
the stochastic equations for the longitudinal 
fluctuations  become 
\begin{align}
m \dot{v}^\pm =& \frac{e^2}{12\pi} (\ddot{v}^{\pm} \pm a \dot{v}^\pm - a^2 v^{\pm})
 + \frac{a e^2}{12\pi} (a v^\mp \mp \dot{v}^{\mp}  ) 
 + eQ{\cal O} \phi
\end{align}
where 
\begin{equation}
{\cal O} = a + e^{a\tau} \partial_{+} - e^{-a\tau} \partial_{-} .
\end{equation}
Here ${\cal O}(\tau)$ acts on $\phi(z(\tau))$ 
and  $\partial_{\pm} = \tfrac{1}{2}( \partial_0 \pm \partial_x )$.

The set of equations can be solved by using the Fourier transformations
\begin{align}
 v^\pm(\tau) &= \int \frac{dw}{2\pi} e^{-i \omega \tau} \tilde{v}^\pm(\omega),
 \\ 
 {\cal O}\phi &= 
\int \frac{dw}{2\pi} e^{-i \omega \tau} {\cal O}\varphi(\omega)
\end{align}
as
\begin{align}
 \tilde{v}^\pm (\omega) = \frac{12\pi}{\omega(e^2 \omega - 12i \pi m)} eQ {\cal O} \varphi(\omega).
\end{align}
Since $(v^+ - v^-)$ is not affected by the quantum field $\varphi$, we can 
safely put it at zero. 
It is consistent with the gauge condition 
$\dot{z} \cdot \dot{z} = 1$, i.e. $\dot{z}_{cl} \cdot \delta \dot{z}=0.$

By using the relation
\begin{align}
 v^\pm = \delta \dot{\xi}+a\delta \tilde{\tau}
  \pm (\delta \dot{\tilde{\tau}} + a\delta \xi)
\end{align}
we can obtain fluctuations for 
 $\tilde{\tau}$ and $\xi$ as
\begin{align}
 \delta \xi (\omega) &= \frac{i}{a^2+\omega^2}
    \frac{12\pi}{(e^2 \omega - 12i \pi m)} eQ {\cal O} \varphi(\omega) \\
 \delta \tilde{\tau} (\omega) &= \frac{a}{a^2+\omega^2}
    \frac{12\pi}{\omega(e^2 \omega - 12i \pi m)} eQ {\cal O} \varphi(\omega) .
\end{align}
The vacuum noise fluctuations for 
${\cal O}\varphi$ can be  calculated as
\begin{equation}
\langle 
 {\cal O}\phi(\tau) {\cal O}\phi(\tau')
\rangle
=\frac{a^4}{32\pi^2} 
  \frac{1}{\sinh^4 \left(\frac{a(\tau-\tau'-i \epsilon)}{2} \right)}.
\end{equation}
It is interesting that this is exactly the same as 
the noise correlation (\ref{phi-correlation2})
appearing in the  stochastic equation for the  transverse
momenta. However the 2-point function
of $\delta\dot{\xi}$ and $\delta \dot{\tilde{\tau}}$ behave 
differently from
the 2-point function of the transverse momenta.
After symmetrization, we have
\begin{align}
 \langle \delta \dot{\xi}(\tau) \delta \dot{\xi}(\tau^\prime) \rangle_S
  =&  
 6 e^2 Q^2 \int d\omega \, \coth \left(\frac{\pi\omega}{a} \right)
\left( \frac{\omega}{a^2+\omega^2}\right)^2 \frac{\omega^3+\omega a^2}{e^4 \omega^2 + (12\pi m)^2},
 \\
 \langle \delta \dot{\tilde{\tau}}(\tau) \delta
  \dot{\tilde{\tau}}(\tau^\prime) \rangle_S 
=&
  6e^2 Q^2 \int d\omega \, \coth \left(\frac{\pi\omega}{a} \right)
\left( \frac{a}{a^2+\omega^2}\right)^2 \frac{\omega^3+\omega a^2}{e^4 \omega^2 + (12\pi m)^2}.
\end{align}
The structure of the integral is quite different from that
appeared in the transverse fluctuations.
The poles at $\pm i \Omega_+$ are the same,
 but the poles at $\pm i \Omega_-$ in the transverse case
are replaced by  poles at $\pm i a$.
Since $a \gg \Omega_-$, the longitudinal fluctuations are affected by
higher frequency modes of the quantum vacuum fluctuations.
Hence,the positions of the poles are of the same order as $a$ 
and we cannot use the derivative expansion with respect to $\omega/a$, which
was used in the case of the transverse fluctuations.
Because of this, we do not know yet the validity of the integrals
and an appropriate way to evaluate them.
This technical problem will be related to 
another problem mentioned in the discussions that
it is not clear what kind of quantities are appropriate to 
describe the thermalization in the longitudinal direction.
We leave further analysis for  future investigations.



\begin{thebibliography}{99} 
\bibitem{ChenTajima}
  P.~Chen and T.~Tajima,
  ``Testing Unruh radiation with ultra-intense lasers,''
  Phys.\ Rev.\ Lett.\  {\bf 83} (1999) 256.

\bibitem{ELI}
  P.G.~Thirolf, D.~Habs, A.~Henig, D.~Jung, D.~Kiefer, C.~Lang, J.~Schreiber, C.~Maia, G.~Schaller, R.~Schutzhold, and T.~Tajima
  ``Signatures of the Unruh effect via high-power, short-pulse lasers,''
  Eur.\ Phys.\ J.\ D 55, 379-389 (2009).

\bibitem{ELIhp}
http://www.extreme-light-infrastructure.eu/

\bibitem{Sciama}
  D.~J.~Raine, D.~W.~Sciama and P.~G.~Grove,
  ``Does a uniformly accelerated quantum oscillator radiate?,''
  Proc.\ R.\ Soc.\ Lond.\ A (1991) 435, 205-215

\bibitem{RavalHuAnglin}
  A.~Raval, B.~L.~Hu and J.~Anglin,
  ``Stochastic Theory of Accelerated Detectors in a Quantum Field,''
  Phys.\ Rev.\  D {\bf 53} (1996) 7003
  [arXiv:gr-qc/9510002].

\bibitem{HawkingRadiation}
  S.~W.~Hawking,
  ``Particle Creation By Black Holes,''
  Commun.\ Math.\ Phys.\  {\bf 43} (1975) 199
  [Erratum-ibid.\  {\bf 46} (1976) 206].

\bibitem{BHthermo}
  J.~D.~Bekenstein,
  ``Black holes and entropy,''
  Phys.\ Rev.\  D {\bf 7} (1973) 2333.
  J.~D.~Bekenstein,
  ``Generalized second law of thermodynamics in black hole physics,''
  Phys.\ Rev.\  D {\bf 9} (1974) 3292.
  J.~M.~Bardeen, B.~Carter and S.~W.~Hawking,
  ``The Four laws of black hole mechanics,''
  Commun.\ Math.\ Phys.\  {\bf 31} (1973) 161.
  S.~W.~Hawking,
  ``Black Holes And Thermodynamics,''
  Phys.\ Rev.\  D {\bf 13} (1976) 191.
 
\bibitem{David:2002wn}
 A.~Strominger and C.~Vafa,
  ``Microscopic Origin of the Bekenstein-Hawking Entropy,''
  Phys.\ Lett.\  B {\bf 379} (1996) 99
  [arXiv:hep-th/9601029];
  J.~R.~David, G.~Mandal and S.~R.~Wadia,
  ``Microscopic formulation of black holes in string theory,''
  Phys.\ Rept.\  {\bf 369} (2002) 549
  [arXiv:hep-th/0203048].
 
\bibitem{spinfoam}
See, for example, 
J.~C.~Baez,
  ``An introduction to spin foam models of BF theory and quantum gravity,''
  Lect.\ Notes Phys.\  {\bf 543}, 25 (2000)
  [arXiv:gr-qc/9905087].

\bibitem{Unruh}
   W.~G.~Unruh,
  ``Notes on black hole evaporation,''
  Phys.\ Rev.\  D {\bf 14}, 870 (1976).

\bibitem{RefEmil}
  E.~T.~Akhmedov, D.~Singleton,
  ``On the relation between Unruh and Sokolov-Ternov effects,''
  Int.\ J.\ Mod.\ Phys.\  {\bf A22 } (2007)  4797-4823.
  [hep-ph/0610391];
  E.~T.~Akhmedov, D.~Singleton,
  ``On the physical meaning of the Unruh effect,''
  Pisma Zh.\ Eksp.\ Teor.\ Fiz.\  {\bf 86 } (2007)  702-706.
  [arXiv:0705.2525 [hep-th]].


\bibitem{Unruhrev}
  L.~C.~B.~Crispino, A.~Higuchi and G.~E.~A.~Matsas,
  ``The Unruh effect and its applications,''
  Rev.\ Mod.\ Phys.\  {\bf 80} (2008) 787
  [arXiv:0710.5373 [gr-qc]].
  CITATION = RMPHA,80,787;

\bibitem{Jacobson}
  T.~Jacobson,
  ``Thermodynamics of space-time: The Einstein equation of state,''
  Phys.\ Rev.\ Lett.\  {\bf 75} (1995) 1260
  [arXiv:gr-qc/9504004].
  CITATION = PRLTA,75,1260;
 
\bibitem{RefUnruhSingular}
  W.~G.~Unruh,
  ``Thermal Bath And Decoherence Of Rindler Space-Times,''
  Phys.\ Rev.\  D {\bf 46} (1992) 3271.

\bibitem{HuLin}
  S.~Y.~Lin and B.~L.~Hu,
  ``Accelerated detector - quantum field correlations: From vacuum
  fluctuations to radiation flux,''
  Phys.\ Rev.\  D {\bf 73} (2006) 124018
  [arXiv:gr-qc/0507054].

\bibitem{Grove:1986}
  P.~G.~Grove,
  ``On An Inertial Observer's Interpretation Of The Detection Of Radiation By Linearly Accelerated Particle Detectors,''
  Class.\ Quant.\ Grav.\  {\bf 3}, 801-809 (1986).

\bibitem{Massar:2005vg}
  S.~Massar, R.~Parentani, R.~Brout,
  ``On the Problem of the uniformly accelerated oscillator: d Jul 1992,''
  Class.\ Quant.\ Grav.\  {\bf 10}, 385-395 (1993).

\bibitem{Brout:1995rd}
  R.~Brout, S.~Massar, R.~Parentani, P.~.Spindel,
  ``A Primer for black hole quantum physics,''
  Phys.\ Rept.\  {\bf 260}, 329-454 (1995).
  [arXiv:0710.4345 [gr-qc]].
  R.~Parentani,
  Nucl.\ Phys.\  {\bf B454}, 227-249 (1995).
  [gr-qc/9502030].
  R.~Parentani,
  Nucl.\ Phys.\  {\bf B465}, 175-214 (1996).
  [hep-th/9509104].

\bibitem{JohnsonHu}  
  P.~R.~Johnson and B.~L.~Hu,
    ``Stochastic theory of relativistic particles moving in a quantum field. I:
  Influence functional and Langevin equation,''
  arXiv:quant-ph/0012137.
  ``Stochastic theory of relativistic particles moving in a quantum field.  II:
  Scalar Abraham-Lorentz-Dirac-Langevin equation, radiation reaction  and
  vacuum fluctuations,''
  Phys.\ Rev.\  D {\bf 65} (2002) 065015
  [arXiv:quant-ph/0101001].
   P.~R.~Johnson and B.~L.~Hu,
  ``Uniformly accelerated charge in a quantum field: From radiation  reaction
  to Unruh effect,''
  Found.\ Phys.\  {\bf 35}, 1117 (2005)
  [arXiv:gr-qc/0501029].

\bibitem{ALD}
M. Abraham and R. Becker, Electricity and Magnetism (Blackie, London, 1937); H.A. Lorentz, The
Theory of Electrons (Dover, New York, 1952), pp. 49 and 253; P.A.M. Dirac, Proc. R. Soc. London A
167, 148 (1938).

\end{thebibliography}
\end{document}